\documentclass[fleqn,usenatbib]{mnras}

\usepackage{newtxtext,newtxmath}

\usepackage[T1]{fontenc}
\usepackage{ae,aecompl}

\usepackage{graphicx}	
\usepackage{amsmath}	
\usepackage{amssymb}	
\usepackage{amsfonts}
\usepackage{epsfig,rotating}
\usepackage{subfig}

\newcommand\mj{{M_{\rm J}}}

\def\av#1{\textcolor{blue}{#1}}

\title[The survival of GI gas giants]{Constraining the initial planetary population in the gravitational instability model}

\author[]{J. Humphries$^1$\thanks{rjh73@le.ac.uk}, A. Vazan$^{2,3}$, M. Bonavita$^4$, R. Helled$^3$  \& S. Nayakshin$^1$
\\
$^{1}$
Department of Physics and Astronomy, University of Leicester, Leicester, LE1 7RH, United Kingdom \\
$^2$
Racah Institute of Physics, The Hebrew University,  Jerusalem 91904, Israel \\ 
$^{3}$
Institute for Computational Science, Center for Theoretical Astrophysics and Cosmology, University of Z{\"u}rich, 8057 Z{\"u}rich, Switzerland\\
$^{4}$
Institute for Astronomy, Royal Observatory, University of Edinburgh, Edinburgh, EH9 3FD, United Kingdom 
}
\date{Accepted XXX. Received YYY; in original form ZZZ}

\pubyear{2017}

\hypersetup{draft}
\begin{document}
\label{firstpage}
\pagerange{\pageref{firstpage}--\pageref{lastpage}}
\maketitle

\begin{abstract}
Direct imaging (DI) surveys suggest that gas giants beyond 20 AU are rare around FGK stars. However, it is not clear what this means for the formation frequency of Gravitational Instability (GI) protoplanets due to uncertainties in gap opening and migration efficiency.
Here we combine state-of-the-art calculations of homogeneous planet contraction with a population synthesis code. We find DI constraints to be satisfied if protoplanet formation by GI occurs in tens of percent of systems if protoplanets `super migrate' to small separations. In contrast, GI may occur in only a few percent of systems if protoplanets remain stranded at wide orbits because their migration is `quenched' by efficient gap opening. We then use the frequency of massive giants in radial velocity surveys inside 5 AU to break this degeneracy - observations recently showed that this population does not correlate with the host star metallicity and is therefore suspected to have formed via GI followed by inward migration. We find that only the super-migration scenario can sufficiently explain this population whilst simultaneously satisfying the DI constraints and producing the right mass spectrum of planets inside 5 AU. 
If massive gas-giants inside 5 AU formed via GI, then our models imply that migration must be efficient and that the formation of GI protoplanets occurs in at least a tens of percent of systems.
\end{abstract}




\section{Introduction}

Two competing scenarios, Core Accretion \citep[CA;][]{PollackEtal96,IdaLin04a,MordasiniEtal12} and Gravitational Instability \citep[GI;][]{Kuiper51b,Boss97,KratterL16}, vie to explain the formation of gas giant planets \citep[for a review of advantages and disadvantages of the theories see][]{HelledEtalPP62014}.
One promising way of separating the two theories is by looking at correlations of the occurrence rate of gas giants with the metallicity of their host stars. CA predicts a strong positive correlation for the occurrence rate of gas giants with host star metallicity since the chances of making massive solid cores is greater in high metallicity discs \citep{IdaLin04b,MordasiniEtal09b, MordasiniEtal12}.

Different predictions were made for GI, and at the moment the dependence on metallicity is unsolved. Several studies suggested that planet contraction times are longer in higher metallicity environments since dust opacities are higher. As planets migrate closer to the star, they are disrupted by stellar tides \citep{BoleyEtal10,Nayakshin10c} unless they contract to much higher densities faster than they migrate. Therefore, fewer GI planets should survive the rapid migration phase around high metallicity hosts, unless grain growth and settling, which could change the trend, is considered \citep{HelledBodenheimer11}.
\cite{FischerValenti05} established that the gas giant occurrence rate correlates positively with host star metallicity, supporting the CA prediction. \cite{MordasiniEtal12} predicted that brown dwarfs made by CA will follow a yet stronger positive metallicity correlation. However, \cite{TroupEtal16} found no significant metallicity correlation for brown dwarfs, confirming earlier suggestions by \cite{RaghavanEtal10}. \cite{MoeEtal18} in fact found that the fraction of close stellar binaries anti-correlates with the host star metallicity. These results indicate that there may be a critical {\em mass} of the secondary below which the secondaries form primarily via CA, and by GI above it. 
Recent observations further support these ideas. \cite{SantosEtal17} found no strong metallicity correlation for planets more massive than $4 \mj$ and a break in the mass function of the planets at around the same mass. The exact mass at which the metallicity correlation reverses is currently disputed with values ranging from 2 $M_J$ \citep{MaldonadoEtal19}, 4 $M_J$ \citep{SantosEtal17,NarangEtal18}, 10 $M_J$ \citep{Adibekyan19,Schlaufman18} to 25 $M_J$ \citep{GodaMatsuo19} for G type stars.
This new sample of observational work demands more thorough theoretical studies of the GI planet formation model in order to understand the origin of the transition mass and the possibility of overlap between the CA and GI planet populations.

In the GI theory, initially massive circumstellar discs fragment at distances of tens to hundreds of AU onto self-gravitating protoplanets within their first $10^4-10^5$ years of life \citep{Rice05, Rafikov05, HallEtal19}. 
These low density protoplanets migrate inwards due to tidal torques from the disc \citep{Tanaka02}.
Depending on the efficiently of gap opening, GI planets may either be stranded in the outer disc or reach separations of less than 10 AU within $\sim10^4$ years \citep{VB06,BoleyDurisen10,BaruteauEtal11}. During this process they cool, contract and eventually collapse due to the dissociation of molecular Hydrogen to form high density, post-collapse `hot start' planets \citep{HB10}.
However, if migration is rapid they may be tidally disrupted before they can reach this collapse point \citep{Boley09,Vorobyov12}. After disruption, fragments release any proto-cores and debris formed by grain growth and sedimentation \citep{Kuiper51b,McCreaWilliams65,Boss97,HelledEtal08,Nayakshin10a,Nayakshin10b,Nayakshin18} back into the disc \citep{NayakshinCha12}.

Predicting the observational outcomes of GI is very difficult due to a range of long-standing uncertainties in the model. These include the likelihood of disc fragmentation and the properties of the disc at that point \citep{Gammie01, Rice05, KratterEtal10, KratterL16}, gas accretion rates onto the planets \citep{MercerStam2017, Nayakshin17a}, gap opening conditions for rapidly migrating planets \citep{MalikEtal15,MullerEtal18}, and the effects of  heavy-element accretion (e.g., planetesimals, pebbles) onto the planets and core formation  \citep{HelledEtal08, JohansenLacerda10,HumphriesNayakshin18}.

Direct Imaging (DI) observations that imply that less than $\sim$ 2.1\% of FGK stars host gas giants or brown dwarfs (0.5-75 $M_J$) at radii of more than 20 AU \citep{ViganEtal17}, though there are two different theoretical interpretations of this low occurrence rate. The first, is suggested by \cite{ViganEtal17} and is based on the population synthesis models of \cite{ForganRice13} (FR13). Planets in the FR13 model open gaps very efficiently and so are typically stranded in the outer disc. Therefore planet formation via GI in these models can only occur in a few percent of systems in order to satisfy the DI constraints.

The second interpretation of the low occurrence rate of planetary mass objects on wide orbits relies on the fact that GI planet migration simulations \citep{VB06,BoleyEtal10,MachidaEtal10,ChaNayakshin11a,BaruteauEtal11,MichaelEtal12} find that gap-opening by planets may be difficult for a variety of reasons \citep{MalikEtal15}. \cite{NayakshinFletcher15}(NF15) argued that if gap-opening is inefficient then GI may occur multiple times in all systems \citep[see also][]{MullerEtal18} since inefficient gap-opening allows planets to migrate to very small separations rapidly. This scenario may satisfy the direct imaging (DI) constraints for wide orbit gas giants by `destroying' such objects efficiently.  In this picture, the majority of planets are able to migrate to separations below 20 AU and are frequently tidally disrupted. They may also undergo runaway gas accretion and become stellar mass companions, depending on uncertainties in gas cooling rates within the Hill spheres of these planets. \citep{NayakshinCha13,Stamatellos15,Nayakshin17a,MercerStam2017}. After these processes are taken into account, wide orbit planetary mass companions are expected to be rare, even if they were initially abundant.
Furthermore, recent infrared spectral observations and RV analysis of CI Tau b by \cite{FlaggEtal19} found that it is best modelled as a `hot start' planet of 11.6 $\mj$ with a period of 9.0 days. Since hot start is typically associated with formation through GI, this adds further weight to the idea that {\bf at least some of the} sub 0.1 AU massive giants form through a combination of GI and efficient migration.

We name our two contrasting scenarios for the orbital evolution of GI planets `quenched' and `super' migration for clarity and simplicity of reference. In this paper we seek to quantify these two scenarios further and to test whether either of them can explain planets more massive than $\sim 4\mj$ at separations less than 5 AU whilst still satisfying the constraints on wide-orbit gas giants.

In this paper we construct a simple population synthesis model and attempt to reproduce these observational results. This allow us to understand the relevant physical processes, as well as their associated degeneracies. 
We have improved upon previous population synthesis studies by coupling migration models with high resolution simulations of GI protoplanet contraction \citep{VazanHelled12}. This allows us to investigate how frequently pre-collapse GI protoplanets are tidally destroyed and therefore how well each model can match to observational constraints on the gas giant population.
In this work we neglect both gas and heavy-element  accretion, core formation and planet-planet scattering, the processes that are expected to have a large impact on the observational predictions of the GI model \citep{Nayakshin16a,ForganEtal18}. Nevertheless, this paper represents another step in the direction of a predictive GI population synthesis. We plan to extend this work to include additional physical processes in the near future.

The structure of this paper is as follows. In Section \ref{Section:Methods} we outline our population synthesis setup and discuss our two extreme migration models: quenched and super migration. We also describe our isolated calculations of protoplanet evolution and their integration into the synthesis model. In Section \ref{sec:full_model} we demonstrate that modelling the pre-collapse protoplanet radius decreases the survival chances for low mass protoplanets and study how this changes for different protoplanet metallicities. In Section \ref{sec:obs_comp} we investigate how observational results constrain our two extreme migration models. Finally, in Sections \ref{Sec:Discussion} and \ref{Sec:Conclusions} we discuss some outstanding issues and list the conclusions of our paper.

\section{Methods}
\label{Section:Methods}

\subsection{Outline of the population synthesis code}

We employ a simple power-law model for the protoplanetary disc that depletes its mass on an exponential timescale. A GI protoplanet is injected into the disc at $t=0$ at a wide separation. Analytic prescriptions are then used to track the planetary migration and determine when it will open a gap in the disc. This disc-planet interaction model is combined with pre-computed models of protoplanet evolution and internal structure \citep{VazanHelled12} to predict whether the GI protoplanet will be tidally disrupted whilst in its extended phase or survive by collapsing to form a much denser post-collapse gas giant planet. We focus on how the survival chances of the protoplanets depend on their mass and metallicity, which we assume equal to the host star and disc metallicity. Since some of the parameters of the problem vary between the observed systems, for each disc we independently pick each parameter from within the ranges shown in Table \ref{table:pop_vals}. This is discussed in more detail in the following section.

\subsection{The disc treatment}
\begin{table}
\centering
\begin{tabular}{ c c c c } \
\setlength{\tabcolsep}{0pt}
Parameter              & Minimum         & Maximum         & Distribution\\
\hline
$M_{\rm disc}/M_{*}$   & 0.1             & 0.3             & linear\\ 
$M_{*} [M_{\odot}]$    & 1.0             & -               & -\\ 
$R_{\rm in}$ [AU]      & 0.01            & -               & -\\
$R_{\rm out}$ [AU]     & 80              & 300             & linear\\
$A_0$                  & 0.5             & 1.0             & linear\\
$t_{\rm disp}$ [Years] & 3 $\times 10^5$ & 3 $\times 10^6$ & log\\
$\alpha_0$             & 0.005           & -               & -\\
\hline
\end{tabular}
\caption{The default values for our initial disc population. $M_{\rm disc}/M_*$ is the initial disc to stellar mass ratio; $M_{*}$ is the stellar mass; $R_{\rm in}$ and $R_{\rm out}$ are the inner and outer radii of the disc; $A_0$ is the starting position of the protoplanet as a fraction of $R_{\rm out}$; $t_{\rm disp}$ is the dispersal time of the disc and $\alpha$ is the dimensionless disc viscosity.}
\label{table:pop_vals}
\end{table}

We consider a disc with a power-law surface density profile $\Sigma \propto R^{-1}$ with inner and outer radii $R_{\rm in}$ and $R_{\rm out}$. In all cases we set the initial stellar mass ($M_*$) to be the solar mass ($1 M_{\odot}$) and the disc mass, $M_{\rm disc}$, with the ratio $M_{\rm disc}/M_*$. The total disc mass is evolved according to the disc dispersal timescale $t_{\rm disp}$ such that  $M_{\rm disc}(t) = M_{\rm 0} \exp(-t/t_{\rm disp})$. The stellar mass is increased by this same value such that the total system mass remains constant. 
The disc is heated by irradiation from the central star which sets a radially dependent temperature of $T(R) = 20$~K~$ (100\, {\rm AU}/R)^{1/2}$. We use an ideal equation of state with $\gamma =5/3$ in order to calculate the $H/R$ ratio of the disc using the thin disc approximation $H/R = c_s/v_K$, where $c_s$ and $v_K$ are the isothermal sound speed and the Keplerian velocity, respectively. The disc viscosity parameter $\alpha$ is the sum of the fixed parameter $\alpha_0$ and a self-gravitational component, as discussed in Section \ref{sec:gap_opening} below.
We position a single, embedded GI protoplanet  at a location $A_0 R_{\rm out}$ within the disc to represent its birth location, with $A_0 \le 1$ (see Table \ref{table:pop_vals}). The system is then evolved following the set of equations detailed in the following sections and the final location of the planet is recorded in order to characterise the outcome of the planet formation process.

\subsection{Migration prescriptions}
\begin{figure*}
\includegraphics[width=1.0\columnwidth]{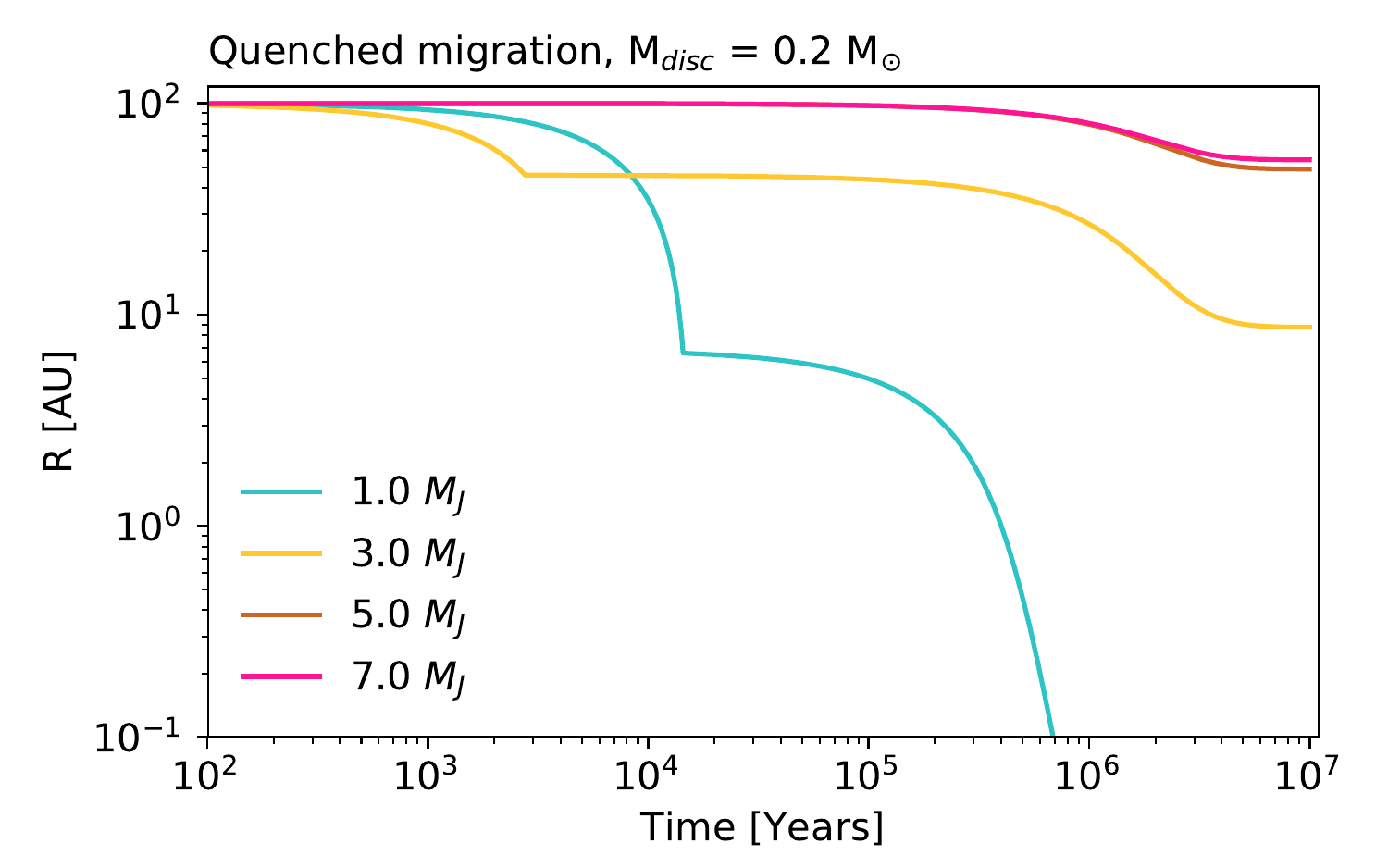}
\includegraphics[width=1.0\columnwidth]{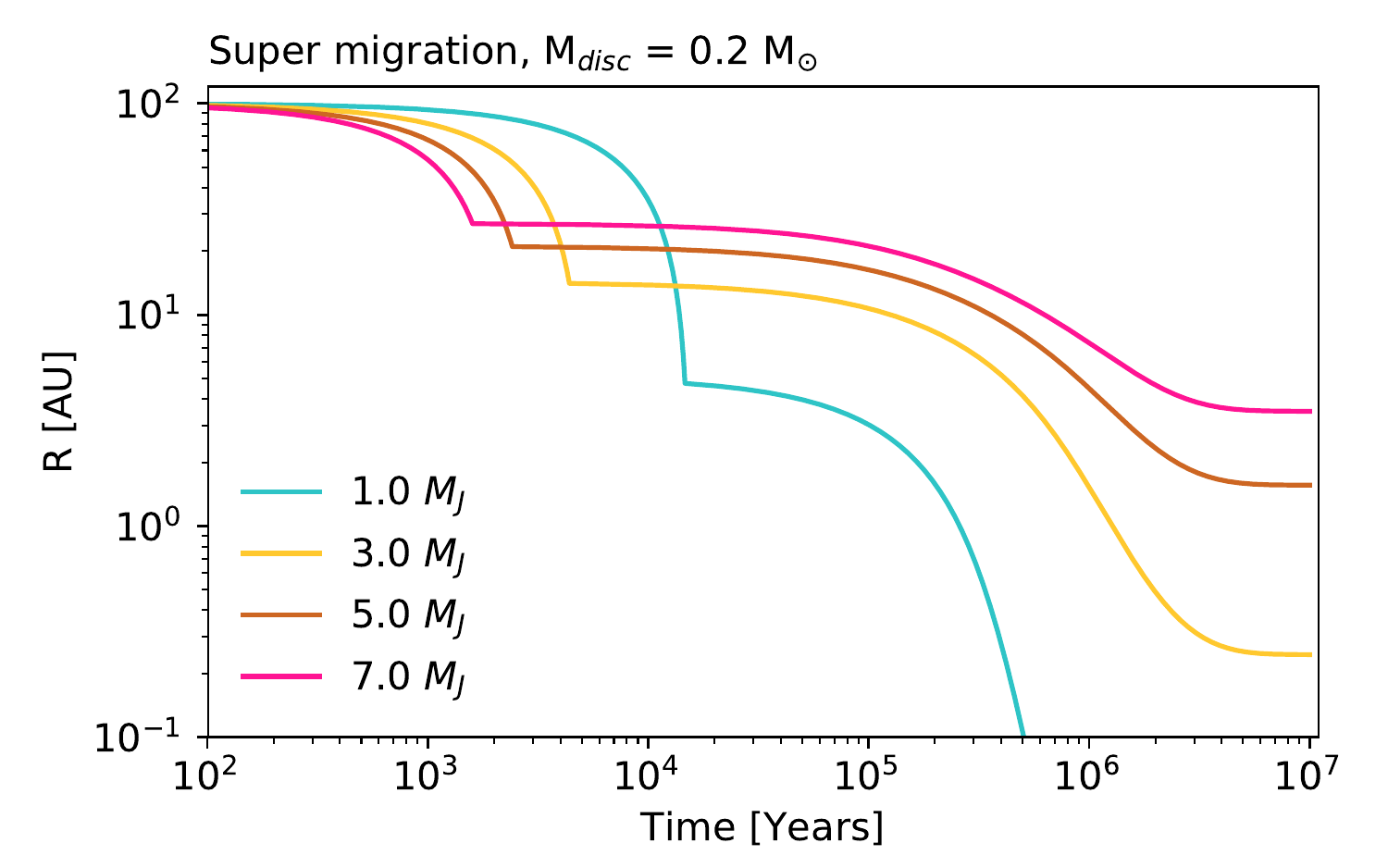}
\caption{Example migration tracks for our quenched and super migration models. In quenched migration, gap opening is efficient and so planets often stall at wide orbits. If gap opening is less efficient, super migration allows planets to migrate to very small separations on short timescales. In both cases the initial type I migration is very rapid due to torques from the massive self-gravitating disc.}
\label{fig:simple_tracks}
\end{figure*}

Planets initially experience type I migration in which they are driven towards the star by the imbalance of torques from the surrounding disc. 
Analytic formulas for type I migration timescales were derived by \cite{Tanaka02} for two and three dimensional isothermal cases,
\begin{equation}
    t_{\rm migI} = \frac{C_{\rm mig}}{\Omega_K} \dfrac{M_*^2}{M_P M_{\rm d}} \left( \dfrac{H}{R} \right)^2,
\end{equation}
where $R$ is the location of the planet and $M_{\rm d} = \pi \Sigma R^2$ is the `local disc mass' (which is different from the total disc mass $M_{\rm disc}$). An alternative prescription outlined in \cite{BaruteauEtal11} is identical, save for a numerical prefactor. The magnitudes of the prefactor $C_{\rm mig}$ for a disc surface density profile of $\Sigma \propto R^{-1}$ are 0.8 and 0.4 for \cite{Tanaka02} in 3D and 2D, and 0.2 for \cite{BaruteauEtal11}. We choose the \cite{Tanaka02} 3D prefactor of 0.8 since it was found to be a good fit in previous 3D simulations \citep{HumphriesNayakshin18, FletcherEtal19}. Additionally, we do not allow the migration timescale to be shorter than the local orbital timescale.

After satisfying the conditions for gap opening (See Section \ref{sec:gap_opening}), planets continue to migrate on the viscous timescale of the disc in the type II regime. 
\begin{equation}
    t_{\rm migII} = \dfrac{1}{\alpha \Omega_K} \left( \dfrac{R}{H} \right)^2 \left( 1 + \dfrac{M_P}{M_{\rm d}} \right).
\end{equation}

There is some uncertainty in this prescription since it does not account for material flowing through the gap and past the planet \citep{DuffelEtal14, DurmannKley2017}. We use it for now in order to remain consistent with previous GI population synthesis work (FR13, NF15, \cite{MullerEtal18}), but see \cite{KanagawaEtal18} for an alternative approach\footnote{our limited tests of the \cite{KanagawaEtal18} migration formulation found that it was a poor fit when compared to rapidly migrating massive gas giants from our previous SPH simulations \citep{FletcherEtal19}.}. Additionally, uncertainties in the type II migration rate are effectively degenerate with uncertainties in the viscous disc $\alpha$ and the efficiency of gap opening. To account for these uncertainties we have chosen to use two extreme migration models which encompass a broad range of possible migration outcomes.

\subsection{Gap opening criteria}\label{sec:gap_criteria}
\label{sec:gap_opening}
The transition between type I and type II migration for giant planets in the outer parts of massive self-gravitating discs is still uncertain, what follows is a brief overview of the topic. 
\cite{LinEtal93} defined a thermal pressure stability criterion, in which planets open a gap instantaneously  if the condition $M_P > 2 M_* (H/R)^3$ is satisfied. \cite{CridaEtal06} later extended this to include a viscous torque criterion for non-migrating planets in which gaps open if 
\begin{equation}
 C_P = \dfrac{3H}{4R_H} + 50 \alpha \left( \dfrac{H}{R} \right)^2 \dfrac{M_*}{M_P} < 1, 
\label{eq:crida}
\end{equation}
where $R_H = R(M_P/3M_*)^{1/3}$ is the Hill sphere of the planet. 
These criteria were developed for stationary planets, yet in reality a rapidly migrating planet may have no time to open a gap before it has migrated across the gap width \citep{MalikEtal15}. 
Additionally, self-gravitating discs have higher levels of turbulence which suggests that their $\alpha$ value should be increased at early times \citep{Rice05}. These two factors complicate the gap opening process for rapidly migrating planets in self-gravitating discs and act to make gap opening less efficient. 

In this work we compare two extreme models for efficient and inefficient gap-opening which we name `quenched' and `super' migration. 
Efficient gap opening leads to quenched migration, planets quickly open gaps and are stranded in the outer disc. We model this by using the Crida gap opening prescription with a fixed value of $\alpha = \alpha_0 = 0.005$. For this relatively small value of $\alpha$, the second (viscous) term in equation \ref{eq:crida} is negligible for the parameter space appropriate to our models, and thus the gap opening behaviour is comparable to that observed in FR13 \footnote{The FR13 study simpl\av{y} used the pressure stability gap opening criterion.}. 

On the other hand, if gravito-turbulence is significant, it is much harder for planets to open gaps. We model this behaviour by following NF15 and modifying $\alpha$ with a self-gravity component such that  
\begin{equation}
\alpha = \alpha_0 + \alpha_{SG} \dfrac{Q_0^2}{Q^2 + Q_0^2}.
\label{eq:alphasg}
\end{equation}
where $Q$ is
\begin{equation}
    Q = \frac{c_s \Omega_K}{ \pi G \Sigma}
\end{equation}
the \cite{Toomre64} parameter, with $Q_{0}$ = 1.4 being the $Q$ below which the disc fragments \citep{BoleyEtal10}. Equation \ref{eq:alphasg} encapsulates an empirical prescription in which the maximum value of the GI-driven turbulent viscosity parameter is $\alpha_{\rm sg}$, and at $Q = Q_0$ we have $\alpha = \alpha_{\rm sg}/2 = 0.06$, consistent with simulations by \cite{Rice05}. As the disc becomes less self-gravitating ($Q$ increases above $Q_0$), self-gravitational turbulence dies away. 
The prescription in equation \ref{eq:alphasg} increases $\alpha$ when the disc is self-gravitating and therefore delays the onset of gap opening due to the appearance of $\alpha$ in Equation \ref{eq:crida}. We term this model super migration since inefficient gap opening allows planets to migrate rapidly from wide orbits down to the inner disc.

\subsection{Example point-mass planet migration tracks}\label{sec:examples}

In this section we assume that the planets are point masses so that they cannot be tidally disrupted. This is done to illustrate the difference in the migration paths of planets in the two extreme migration regimes. Figure \ref{fig:simple_tracks} shows the difference between the quenched and super migration scenarios for 1, 3, 5 \& 7 $M_J$ planets in a disc of initial mass $M_{\rm disc} = 0.2 M_{\odot}$. The left hand panel shows the quenched migration model; these planets open gaps efficiently and so are mostly left stranded in the outer disc, except for the lowest mass planet.
In contrast, gap-opening is inefficient for the super migration models, and so these planets typically migrate to very small separations. 
For both models, more massive planets initially migrate faster since type I migration depends linearly on the planet mass. However, massive planets are more likely to open gaps and so they transition into the slower type II migration regime sooner. This means that massive planets are generally left stranded at wider orbits than their lower mass cousins.

The difference between these two models has led to two different interpretations of the \cite{ViganEtal17} constraint. In the quenched migration model, GI occurs only in 1-10\% of systems since the total number of planets stranded at large radii must be small. In contrast, with super migration GI may happen multiple times in each system. Since gaps open less often, only a relatively small fraction of all planets are actually left in the outer disc, the others rapidly migrate to the inner disc.

\begin{figure}
\includegraphics[width=1.0\columnwidth]{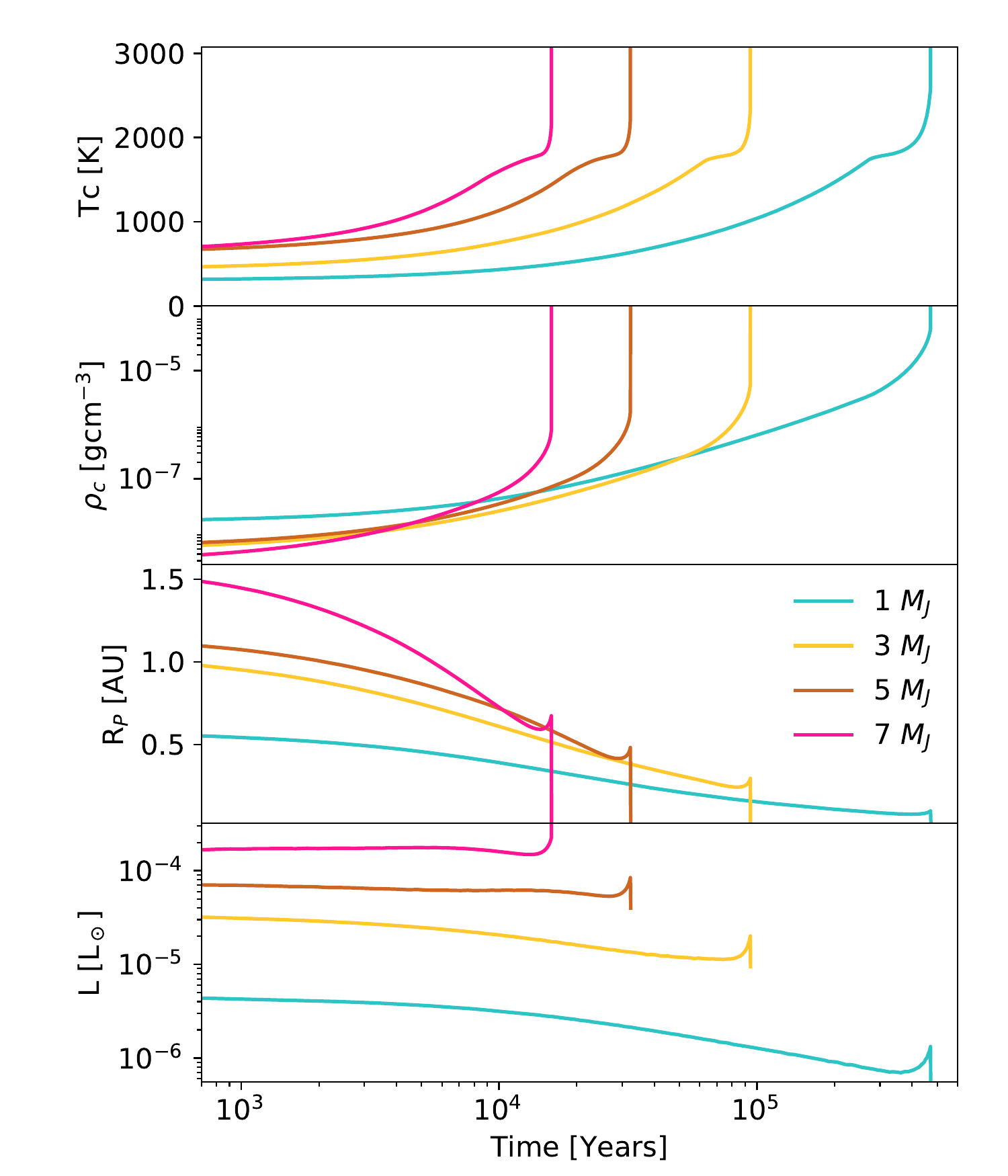}
\caption{Evolution of pre-collapse  gaseous planets, calculated in isolation for 1,3,5 and 7 $M_J$ planets. From top to bottom: Panel 1: central temperature [K]; panel 2: central density [g/cm$^{-3}$]; panel 3: protoplanet radius [AU]; panel 4: protoplanet luminosity [$L_{\odot}$].
Notice that the timescale  for collapse is considerably shorter for higher mass planets. }
\label{fig:allona_models}
\end{figure}

\begin{figure}
\includegraphics[width=1.0\columnwidth]{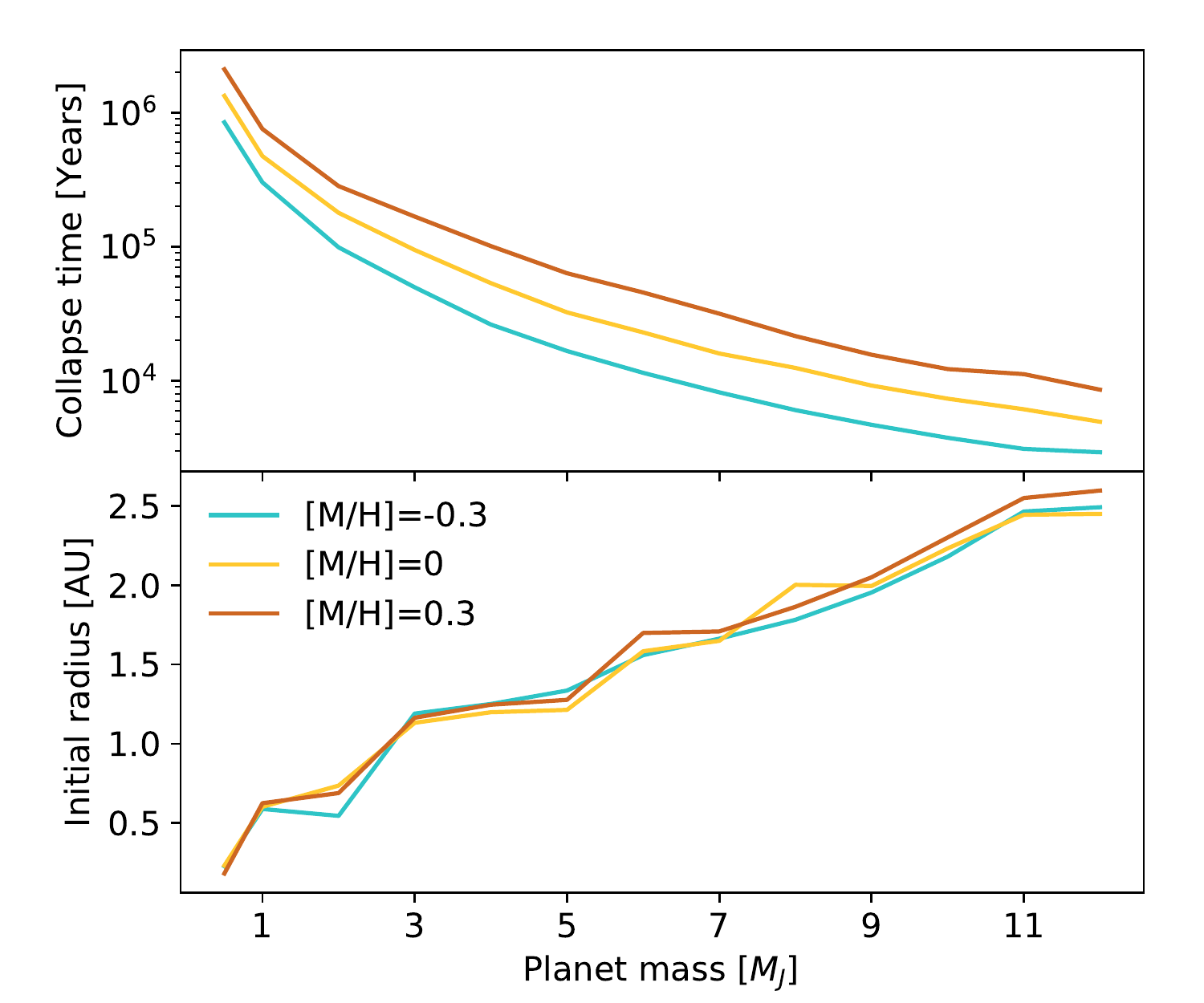}
\caption{Protoplanet collapse time and initial radii as a function of initial planet mass. More massive protoplanets cool and collapse faster due to their higher luminosities. We also show the results for different metallicities; protoplanets with a higher metallic composition are more opaque and therefore cool and collapse slower.}
\label{fig:allona_models2}
\end{figure}

\subsection{Thermal evolution of isolated protoplanets}\label{sec:isolated}

Protoplanets born through GI are initially cold, extended objects with radii $\sim$ 1 AU \citep{HelledEtal06,HelledEtal08}. After a period of contraction, the increasing temperature in their centres causes molecular Hydrogen to disassociate and they undergo collapse to a second core state \citep{Bodenheimer74}, becoming high-density planets with radii of a few $R_J\sim 10^{-3}$~AU. The post collapse configuration is therefore much more stable against tidal disruption. 

In this paper the protoplanet's evolution is modeled by a 1D planetary evolution code that solves the standard stellar structure and evolution equations \citep{HelledEtal06,HelledEtal08,VazanHelled12, VazanEtal13, VazanEtal15}. The code calculates the evolutionary tracks of the protoplanet contraction and collapse, on a fully implicit, adaptive grid numerical scheme. The protoplanet is assumed to consist of hydrogen and helium with a protosolar ratio, using the equation of state of \cite{SaumonEtal95}.  During the cooling and contraction of the protoplanet, heat is transported by convection, conduction or radiation, depending on the local conditions at each time. In radiative regions the opacity is calculated from the \cite{PollackEtal85} ISM opacity table, which is based on the size distribution relevant for interstellar grains. Since the opacity is of great importance for the protoplanet's cooling, we also consider cases with high and low opacity. 
For simplicity, we neglect effects of grain growth and settling, which could modify the contraction timescales  \citep{HelledBodenheimer11}. The initial model of the protoplanet is derived assuming an adiabatic internal structure.

The dynamical collapse of a protoplanet occurs when molecular hydrogen dissociates. As a result, the planetary structure becomes dynamically unstable, i.e., $\int_{0}^{M} (\gamma - 4/3)p\,dV <0$, where $\gamma={\partial \ln p(\rho,s,X)}/{\partial \ln \rho}$
is the adiabatic index.  During the collapse a dynamic (rapid) change is expected, and we use a special routine which yields the outcome of the rapid phase, by making use of the quasi-dynamic algorithm \citep{RakavyEtal67,KovetzEtal09}, then the post-collapse hydrostatic configuration is found. Figure \ref{fig:allona_models} shows the evolution of several of the key outcomes from the evolution models. Generally, the central temperature and density increase as protoplanets cool and collapse, leading to a decrease in the radius and luminosity\footnote{The luminosity decrease is an outcome of the radius decrease during the contraction phase; the luminosity does increase substantially after dynamical collapse}. 

These calculations were performed for three values of metallicity for the planets, [M/H] = -0.3, 0, and 0.3. We make the simplest assumption here that the planet opacity is directly proportional to the metallicity of the planet, and that the latter is equal to the metallicity of the host star. This is clearly an over-simplification  as the opacity can actually decrease with increasing metallicity due to grain growth and settling as discussed in \citet{HelledBodenheimer11}. Figure \ref{fig:allona_models2} shows the collapse times and initial radii for our complete suite of isolated planet calculations. More massive planets typically collapse faster. As expected, when assuming that opacity scales with metallicity, the more opaque (higher metallicity, [M/H]) planets cool and collapse on longer timescales \citep{HB10}.

\subsection{Disc irradiation of protoplanets}
\label{sec:thermal_corr}

Figure \ref{fig:allona_models} shows the evolution of each pre-collapse protoplanet in isolation. In reality, irradiation by an external radiation field can delay the protoplanet's contraction, since the outgoing luminosity is the internal planet luminosity {\em minus} luminosity incident onto the planet. A planet embedded in a disc is essentially `younger' than it would be in isolation.

Our isolated planets evolve according to equation
\begin{equation}
    \frac{dE}{dt} = -L_0(t)\;,
    \label{eq:dE_dt}
\end{equation}
where $E(t)$ and $L_0(t)$ are the total energy and radiative luminosity of the planet, respectively. A planet immersed into the disc with midplane temperature $T_{\rm mid}$ cools according to
\begin{equation}
    \frac{dE}{d\tau} = -L = -L_0(t) + 4\pi R_{\rm p}^2 \sigma_{\rm B} T_{\rm mid}^4\;,
    \label{eq:dE_dtau}
\end{equation}
where $R_{\rm p}$ is the current radius of the planet. Note we introduced delayed time $\tau$. Dividing equation \ref{eq:dE_dt} by equation \ref{eq:dE_dtau} we get,
\begin{equation}
    \frac{d\tau}{dt} = \frac{L}{L_0} = 1 - \frac{4\pi R_{\rm p}^2\sigma_{\rm B}T_{\rm mid}^4}{L_0}
\end{equation}
This equation is integrated to find the function $\tau(t)$. The property $X(t)$ of the planet at any time $t$ is then given by 
\begin{equation}
    X(\tau) = X\left[(\tau(t))\right]
\end{equation}

This treatment is approximate since in reality the outer layers of an irradiated planet adjust to the irradiation it receives, so the planet is expected to have somewhat different temperature and density profiles \citep{VazanHelled12}. However, our simple approach allows us to  adjust the planetary conditions to the changes in the disc temperature, $T_{\rm mid}$, during planet migration, and therefore take 
the "thermal bath" effects of the disc onto the planet approximately \citep{CameronEtal82,VazanHelled12}.

\subsection{Tidal disruption of planets}
\label{sec:disruption}
The exact way in which planets loose mass when they come close to filling their Roche radius $R_{\rm H}$ depends on many factors, such as their equation of state, their spin, and the temperature profile in the outer envelope of the planet \citep[see][for the case of mass transfer in stellar binaries]{Ritter88}. For simplicity, in this paper the planets are assumed to be completely disrupted once their radius exceeds the disruption radius,
\begin{equation}
    R_{\rm disr} = \frac{2}{3} R_{\rm H}\;.
    \label{eq:td0}
\end{equation}
The motivation for this comes from the fact that molecular-hydrogen dominated planets have an adiabatic index $\gamma \approx 7/5$ in a broad range of parameter space. Polytropic spheres with such $\gamma$ migrating through the protoplanetary disc were shown to be unstable to mass loss via Roche lobe overflow: once they start to loose mass, their radius expands whereas $R_{\rm disr}$ contracts, which leads to a runaway -- a nearly dynamical disruption \citep{NayakshinLodato12}. 

The assumption of a complete planet disruption breaks down if a core builds inside the protoplanet \citep{HelledEtal08,HS08,Nayakshin10b}, or if pebble accretion builds an inner metal-rich "fuzzy core" \citep{Nayakshin18}. In this case a partial disruption of the gaseous envelope is possible but since we neglect planetesimal/pebble accretion and core formation in this paper we leave these effects for a future study.

\section{The migration and disruption model}
\label{sec:full_model}
\begin{figure*}
\includegraphics[width=1.0\columnwidth]{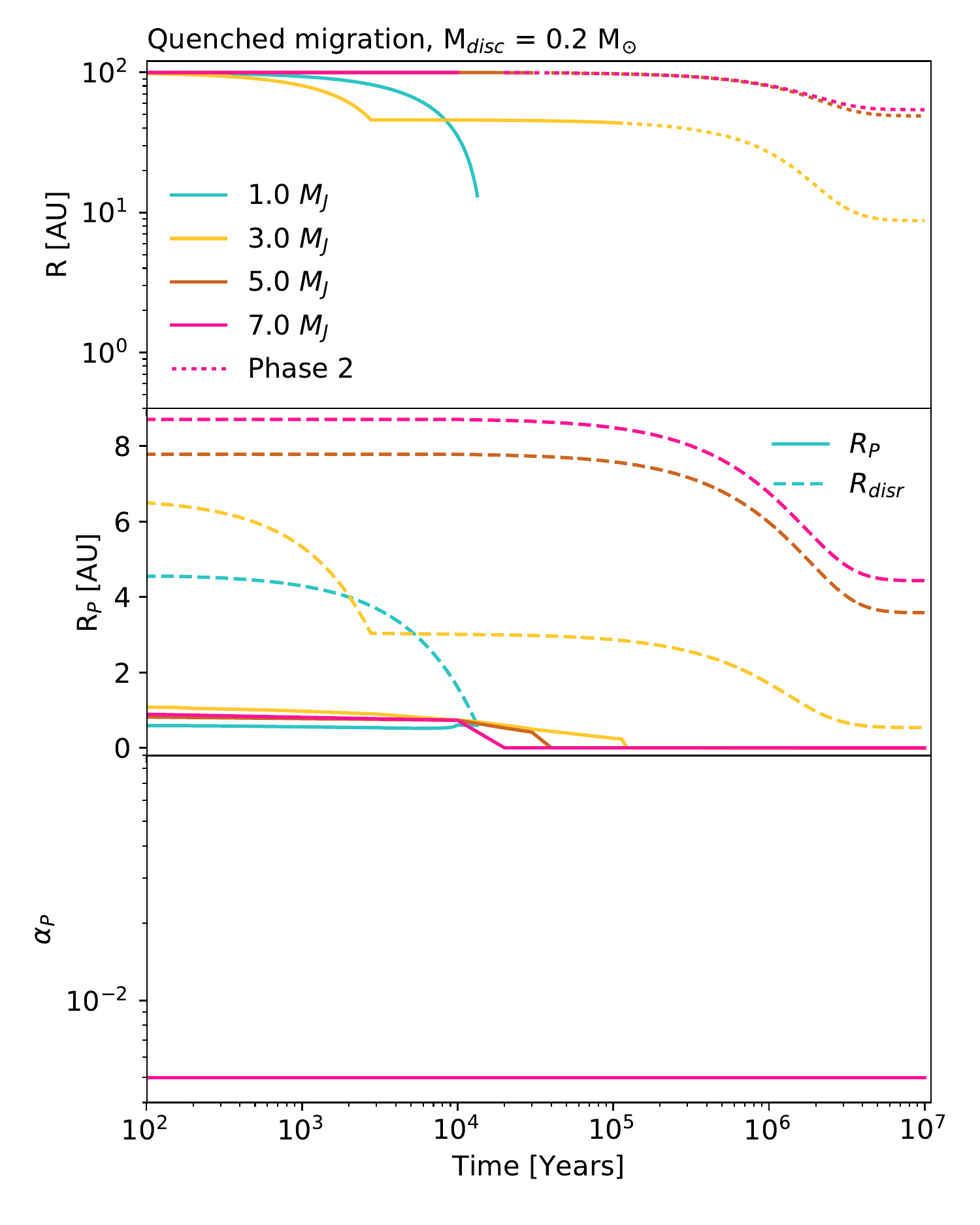}
\includegraphics[width=1.0\columnwidth]{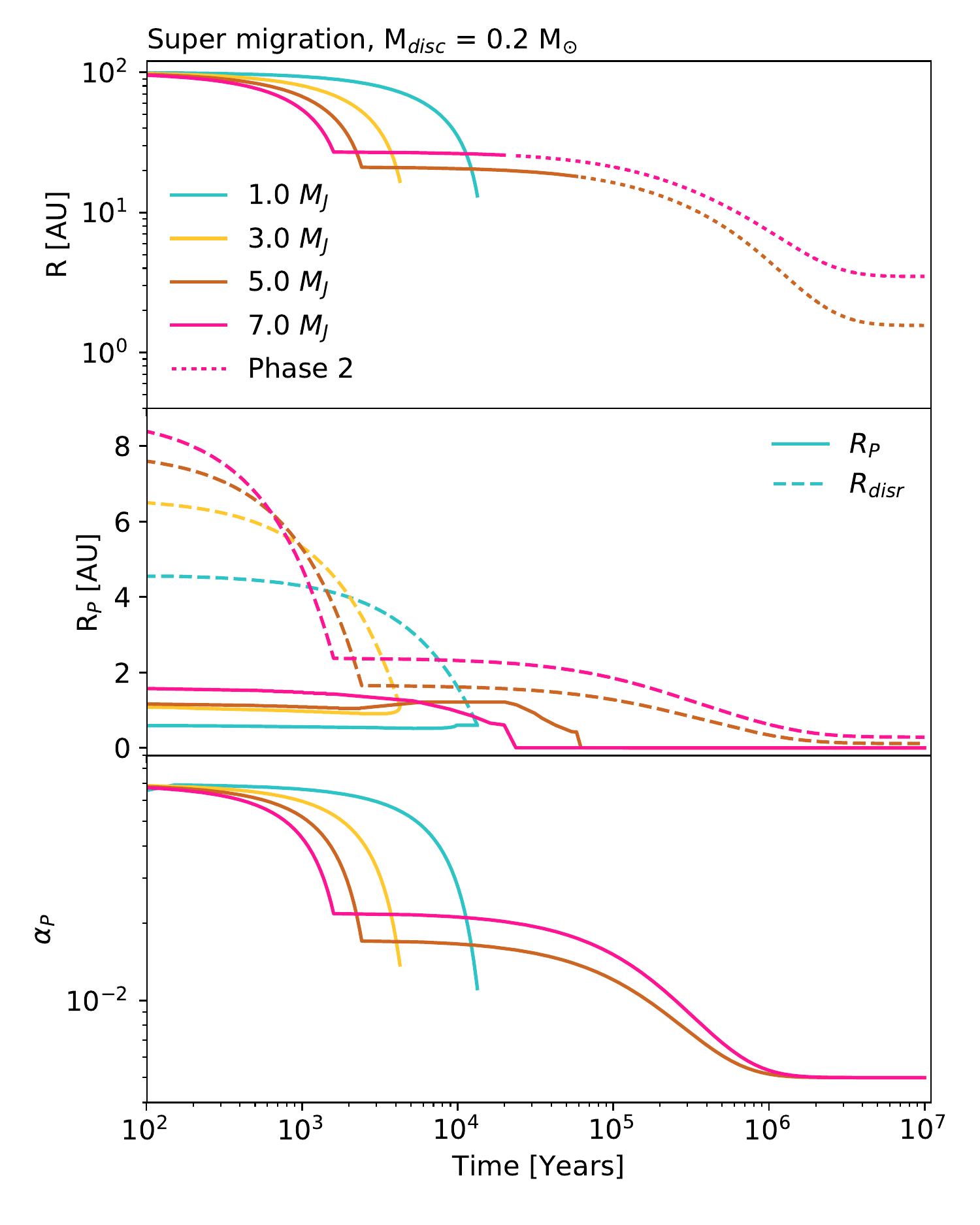}
\caption{Example planet evolution with migration and tidal disruption included for protoplanets of different masses. Top panel: Migration tracks. Those that end abruptly show the planets that were tidally disrupted. Dotted lines represent the evolution of post-collapse planets that survived disruption. Middle panel: Fragment radius (solid), 2/3 Hill sphere (dashed). When these lines meet the fragment is disrupted. After collapse, GI planets typically have radii of $\sim 2R_J$. Bottom panel: $\alpha$ parameter at the location of the planet ($\alpha_P$). Due to a gravito-turbulence enhanced $\alpha$, super-migrating planets avoid stalling due to gap opening for much longer than planets in the quenched migration regime.}
\label{fig:MP_runs}
\end{figure*}

\subsection{Planet evolutionary paths}\label{sec:paths}

We now combine planet contraction and migration in one code to calculate the evolution of GI clumps from their birth to the disc dissipation. The top panel of Figure \ref{fig:MP_runs} shows the migration tracks of the 1, 3, 5, \& 7 $\mj$ planets for the same initial conditions as in Figure \ref{fig:simple_tracks} for both the quenched and the super migration models. In Figure \ref{fig:MP_runs} planet tracks end prematurely if the planet is tidally destroyed. If planets survive tidal disruption by collapsing into the second very dense configuration (see Section \ref{sec:isolated}), then we switch from solid to dotted curves. These planets may survive to the present day (unless they migrate all the way into the star).

We find that planets in the super migration case are much more likely to be tidally disrupted since their migration time is frequently shorter than their collapse time. The middle panel in Figure \ref{fig:MP_runs} shows the planetary radius (solid) and the disruption radius $R_{\rm disr}$ (dashed, equation \ref{eq:td0}); when these two curves meet the planets are tidally disrupted. In general, if protoplanets manage to open gaps whilst in the outer disc then they survive. Planets come much closer to filling their disruption radius in the super migration case.

The bottom panel of Figure \ref{fig:MP_runs} shows the disc $\alpha$ parameter at the location of the planet. In the super-migration scenario the disc viscosity is much larger due to the addition of the gravito-turbulence term $\alpha_{\rm sg}$. This larger viscosity is the reason why the planets in this case migrate much closer in to the star than in the quenched migration case.

\subsection{A grid of models}\label{sec:grid}
\begin{figure*}
\includegraphics[width=1.0\columnwidth]{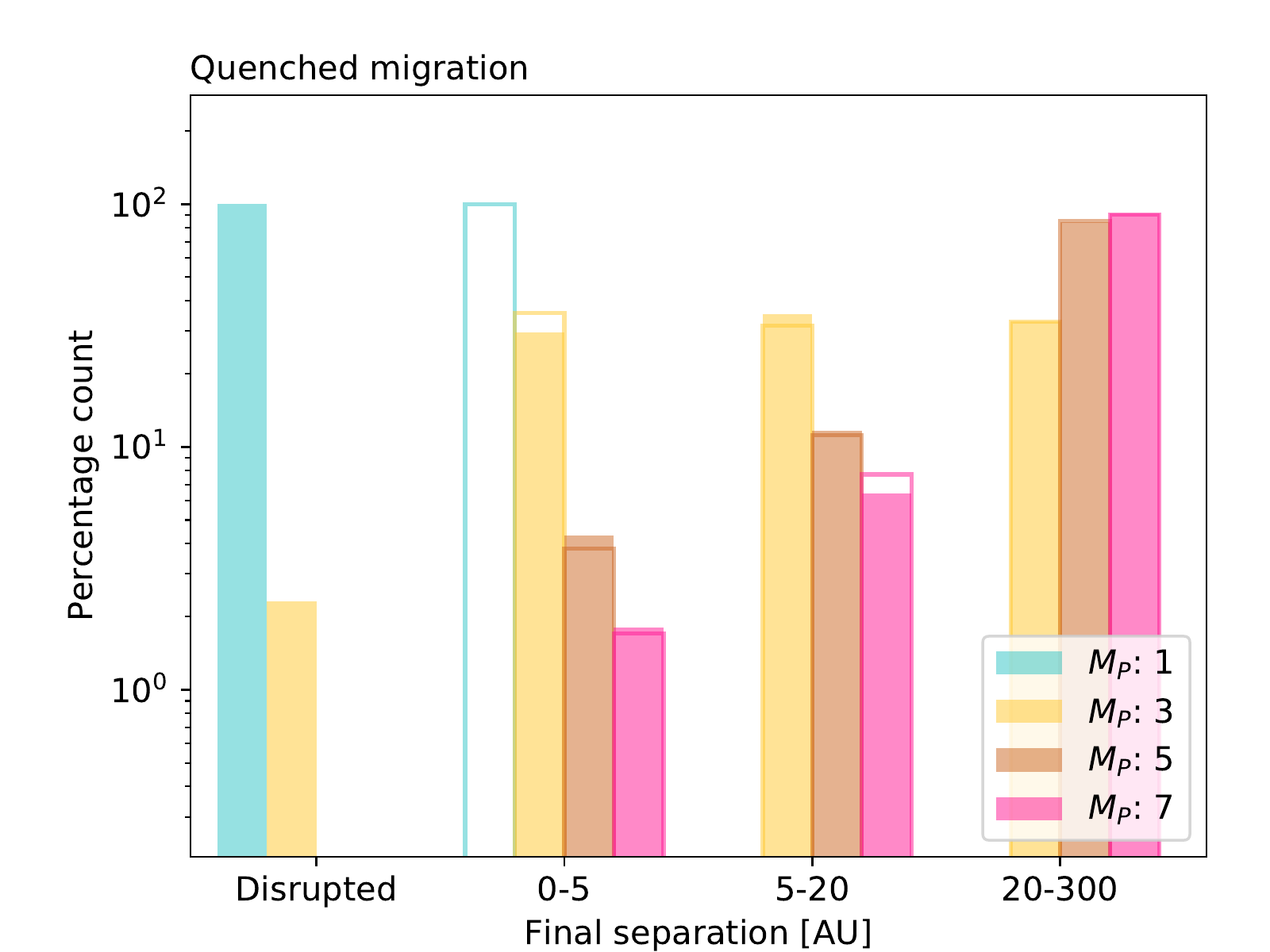}
\includegraphics[width=1.0\columnwidth]{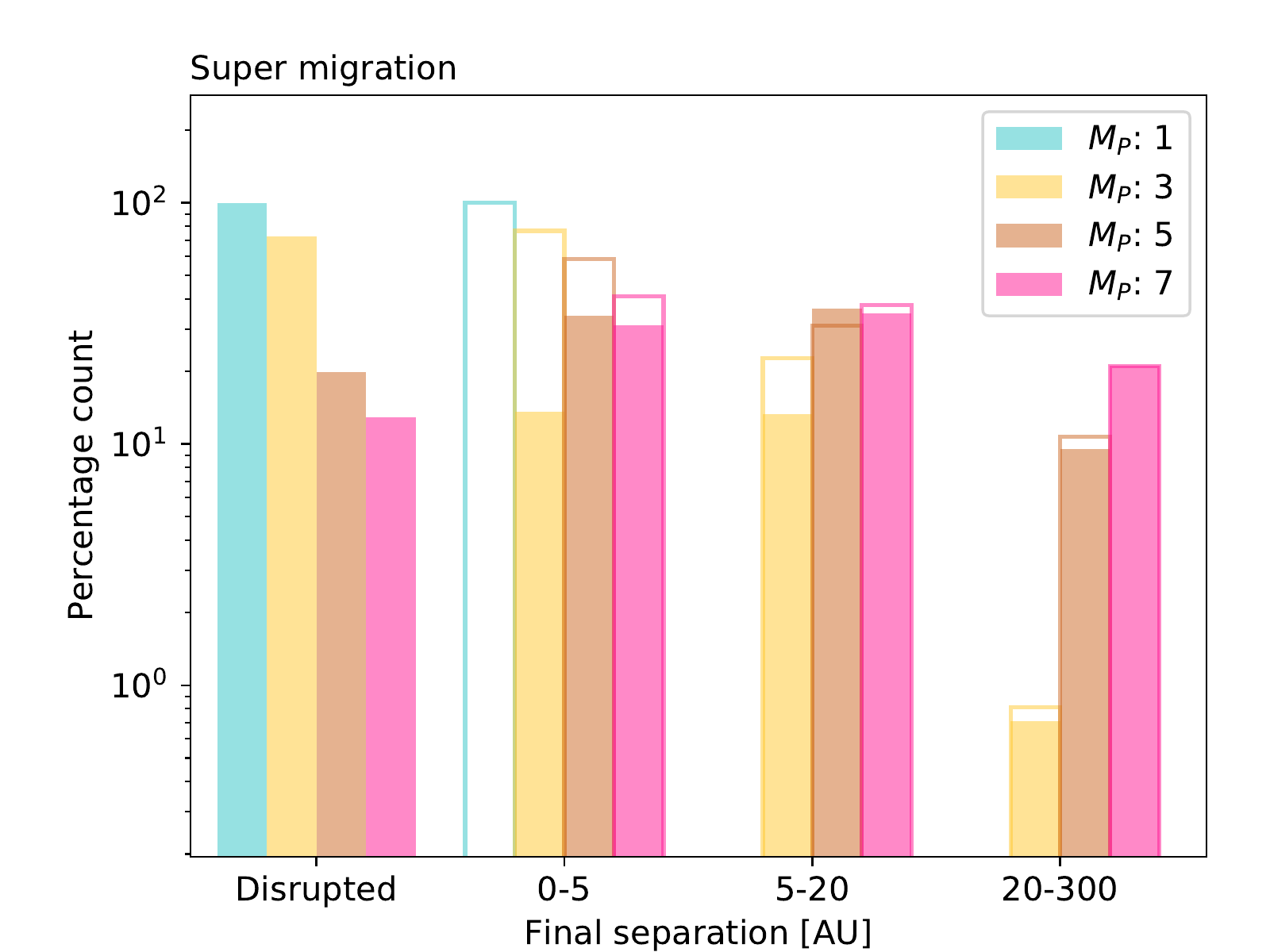}
\caption{The separation distribution of the final population of planets for the quenched and super migration scenarios. Filled in histograms show full models that include migration and tidal disruptions, whereas the outline histograms neglect tidal disruptions, assuming point-mass planets.
The disrupted bin shows the tidally disrupted planets ($R_P > 2/3 R_H$) while planets that reached $R_{in}$ are counted in the 0-5 AU bin. The disc parameters are chosen from the ranges given in Table \ref{table:pop_vals}. Modelling the planetary radius allows us to simulate tidal disruption events, planets are more likely to be tidally disrupted in the super migration model.}
\label{fig:RP_bar}
\end{figure*}

To explore how these effects impact a population of planets in different discs, we repeat the experiments from Figure \ref{fig:MP_runs} for 1000 discs with parameters randomly chosen from the ranges shown in Table \ref{table:pop_vals}. Given the paucity of observational constraints for young discs and the absence of better solutions, we choose each parameter independently from the others. We also repeat these same experiments for point-mass planets (that cannot be tidally disrupted) to emphasize the importance of the detailed modelling of planet radius evolution. 

Figure \ref{fig:RP_bar} shows the outcome of these experiments for the four planetary masses. The four bins along the horizontal axis show the percentage of planets that were tidally disrupted; migrated to a separation inside 5 AU (including those that reached the disc inner boundary $R_{\rm in}$); and stalled between $5-20$ or $20-300$~AU. The outlined bars show the results for point-mass planets whereas the filled-in bars show the full model with tidal disruptions included.

Let us first focus on the simpler point-mass planet case (no tidal disruptions). As expected, we find that planets are much more likely to remain at wide orbits in the quenched migration model. As much as $\sim 30-70$\% of planets remain beyond 20~AU in this model, except for $M_P = 1\mj$ planets which tend to migrate very close to the star. In contrast, for super migration we find only a percent level occurrence rates for planets in the 20-300 AU bin. Most of these planets ($\sim 80-100$\%) end up migrating to separations closer than 20 AU.
The trends seen in Figure \ref{fig:RP_bar} with planet mass are also as expected. Jupiter mass planets tend to be too low mass to open gaps in the outer disc and typically migrate inside 5 AU.  Higher mass planets on the other hand, are more likely to be left in the outer disc due to their increased chance of gap opening.

Shifting our focus to the more self-consistent calculations, we see that tidal disruptions destroy all of the $1\mj$ planets in both quenched and super migration cases. 3 $\mj$ planets are also likely to be disrupted in the super migration model, but remain safe under quenched migration along with the 5 \& 7 $\mj$ planets. Due to their shorter collapse timescales, more massive protoplanets are more likely to reach their post-collapse phase before they can be tidally disrupted. The ensuing tidal disruption occurs at $\sim$ 10-20 AU, since this is the radius at which pre-collapse protoplanets typically fill their Hill radii. This explains why the effect is most significant in the 0-5 AU bin. 
Additionally, heating from the thermal bath acts to delay the collapse of protoplanets and leads to an even higher chance of tidal disruption, in agreement with results of \cite{VazanHelled12}.

\subsection{The importance of gap opening}\label{sec:gap_imp}

\begin{figure*}
\centering
\subfloat[][]{\includegraphics[width=1.\columnwidth]{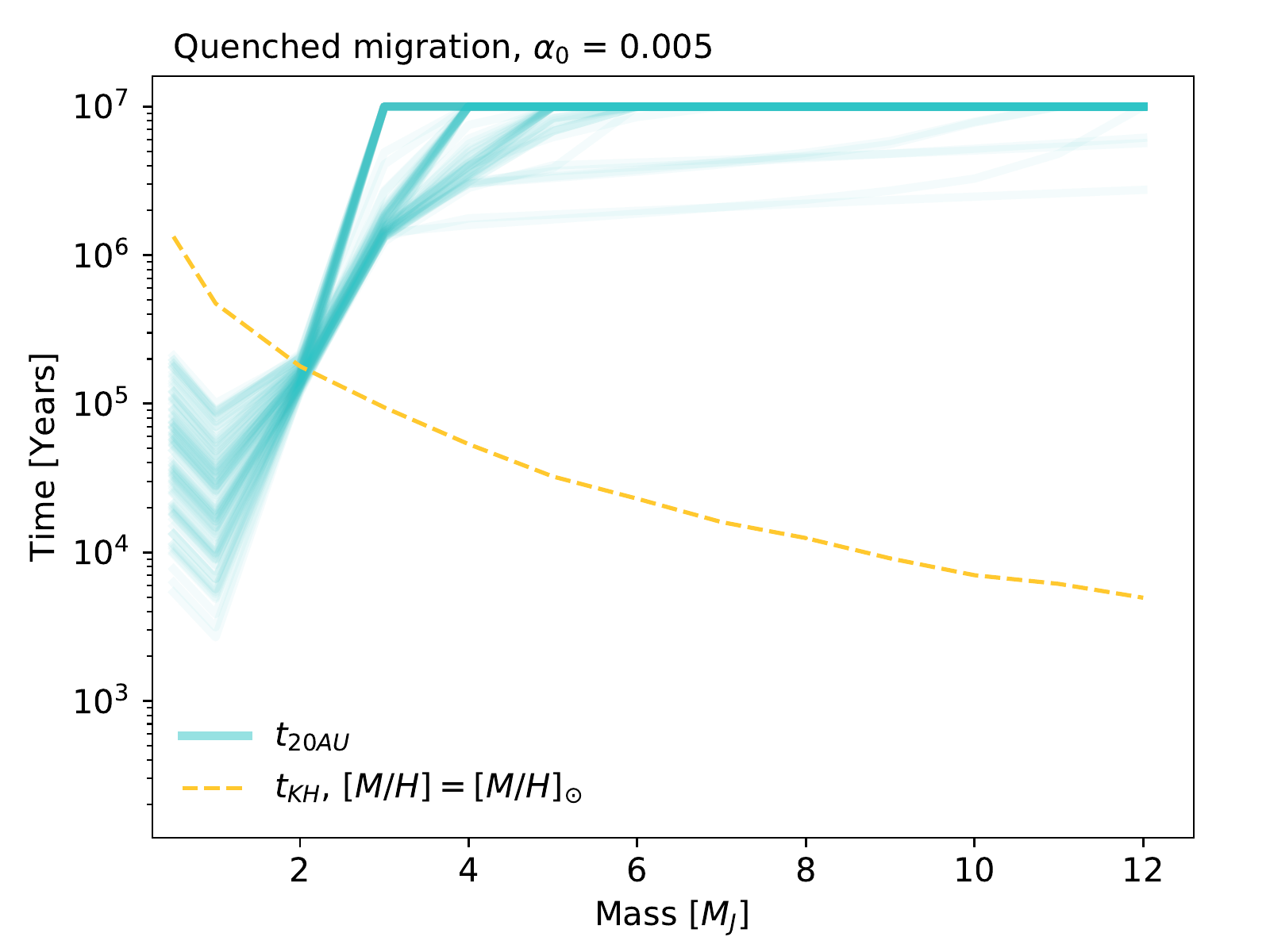}}\quad
\subfloat[][]{\includegraphics[width=1.\columnwidth]{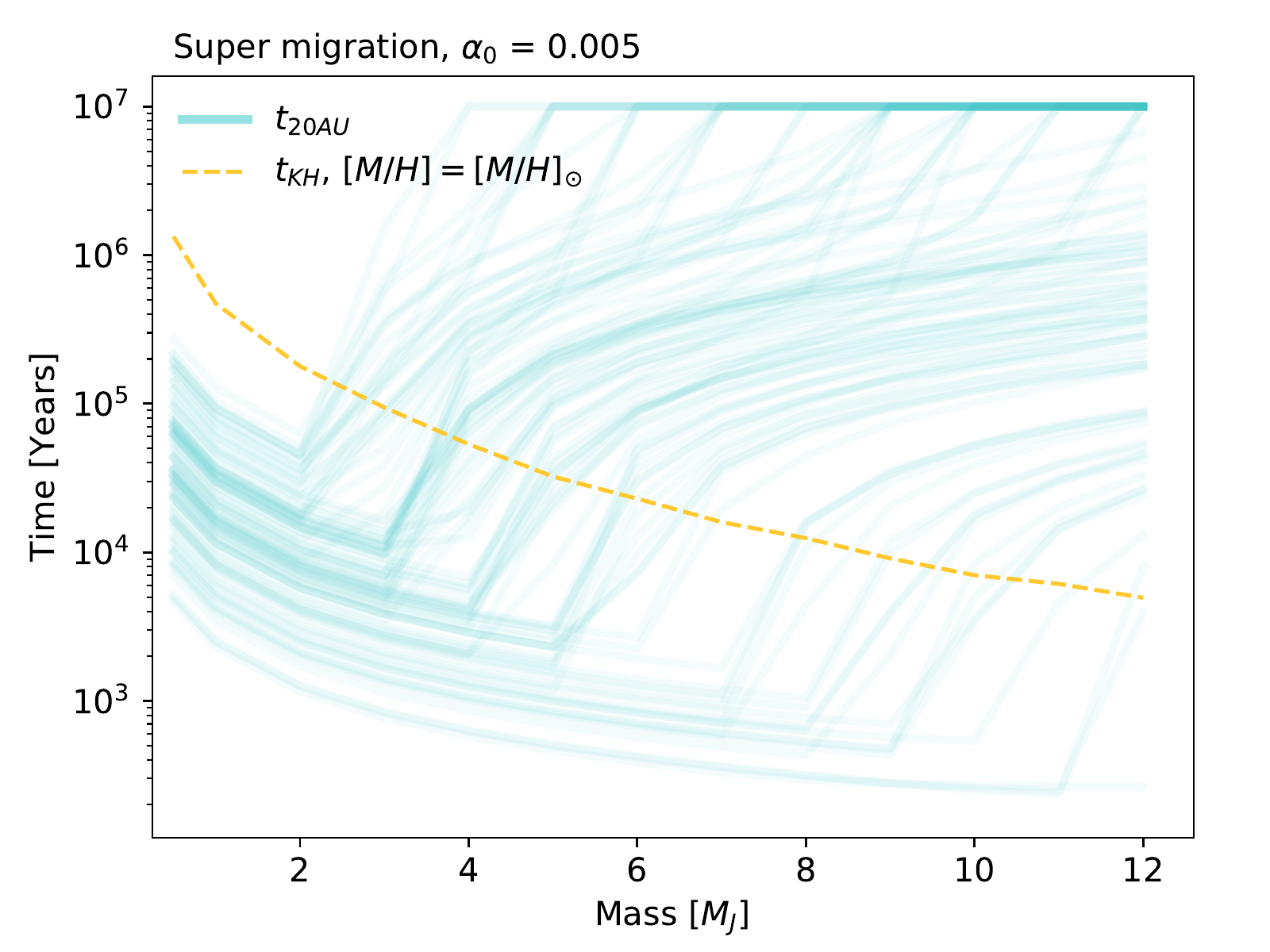}}\\
\caption{The time taken by planets of different mass to migrate to 20 AU (cyan tracks) for the quenched (left) and the super migration (right) models, compared with the  cooling and collapse time scale $t_{\rm KH}$ (yellow dashed line). Different lines represent 100 different initial discs (see Section \ref{sec:full_model} for detail). The migration time increases for planets that open gaps, allowing protoplanets to collapse before they can be tidally disrupted. Planets for which $t_{\rm 20AU} > t_{\rm kh}$ tend to survive after disc dissipation.}
\label{fig:M_transition}
\end{figure*}

We now study this preferential tidal disruption of low mass protoplanets by comparing their collapse timescales and migration histories. We present the {\em point-mass} planet migration experiments described in Section \ref{sec:grid} for a wider selection of planetary masses, that is, $M_P = 0.5, 1, 2, 3, ... 12$~$\mj$. 
Point-mass planets cannot be tidally disrupted, but these experiments allow us to measure their migration times. As we found that tidal disruption of planets in the self-consistent planet contraction models typically takes place at around $10-20$~AU we define the planet migration time as the time it takes the planet to migrate to 20 AU. If the planet never reaches 20 AU then we set the migration time to the maximum disc lifetime of $10^7$ years.

Figure \ref{fig:M_transition} shows these migration times as a function of planet mass for the quenched (left) and the super (right) migration models. Each cyan line represents a single set of initial disc conditions, for each disc we measured the $t_{20AU}$ migration time for every planet mass. The collapse timescale as a function of planet mass at solar metallicity, [M/H] = 0, is plotted with a dashed yellow line.

The lower sweep of the cyan tracks represents planets that reached 20 AU during type I migration. At some point, the tracks abruptly transition to the upper part of the plot which corresponds to gap opening and much slower type II migration.
We see that the type I migration times are always below the yellow curve. These protoplanets would be tidally disrupted if they were not point masses because they have not yet collapsed. The planets on the upper part of the tracks are above the yellow curve; these planets are able to safely cool, collapse, and form post-collapse gas giants. 

The point at which tracks jump from the lower to upper branch indicates the gap opening mass for the corresponding disc. 
We see that under quenched migration only the 1 $M_J$ mass planets reach 20 AU in the type I regime. More massive planets open gaps and take $10^6$ years or more to reach 20 AU and can therefore survive. A much larger fraction of the super migration planets reach 20 AU within $10^5$ years since they are less likely to open gaps. Their migration timescales remain short compared to their collapse timescales and so they are frequently tidally destroyed.

We can see by comparing the lower cyan tracks with the yellow collapse time curve in Figure \ref{fig:M_transition} that type I migration is always faster than the collapse timescale. However, if a protoplanet is able to open a gap in the outer disc then it slows down and has plenty of time to reach the post-collapse phase. This means that GI protoplanets can only survive if they are able to open a gap in the disc beyond $\sim$20 AU. Due to Equation \ref{eq:crida}, more massive planets are more likely to open gaps and therefore are more likely to survive. Equally, gap opening is easier in the quenched migration model and so these planets are safer from tidal disruption. These results neatly explain the bias towards the survival of more massive planets in the quenched migration model seen in Figure \ref{fig:RP_bar}.
We find gap opening masses of 2 $M_J$ and 3 $M_J$ for the quenched and super migration models, though these values are dependent on the selected parameters for the initial disc.

\subsection{Dependence on host star metallicity}

\begin{figure*}
\centering
\subfloat[][]{\includegraphics[width=1.\columnwidth]{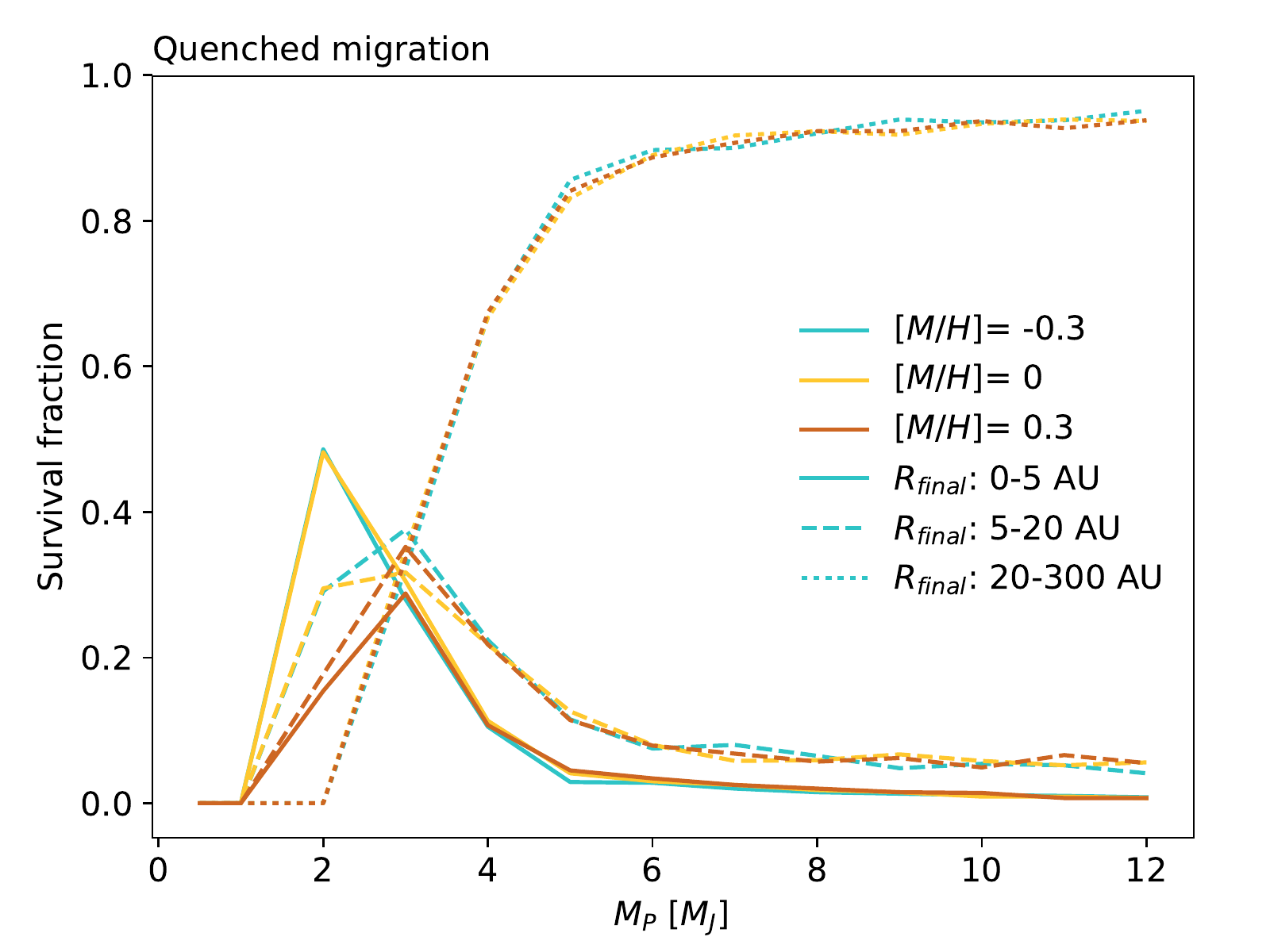}}\quad
\subfloat[][]{\includegraphics[width=1.\columnwidth]{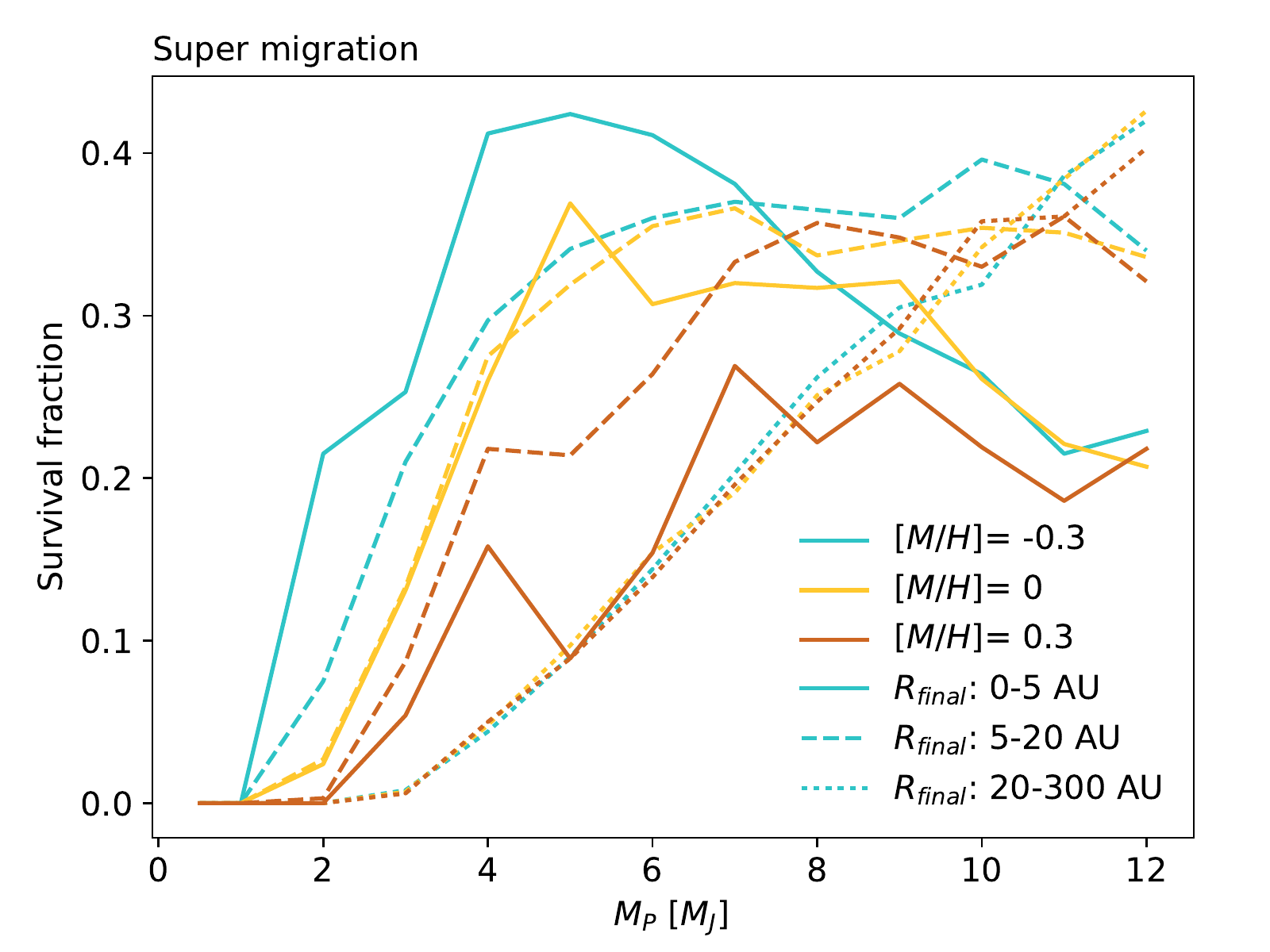}}\\
\caption{Survival percentage as a function of planet mass for the quenched and super migration models. The solid, dashed and dotted lines represent the survival percentage in bins of 0-5, 5-20 \& 20-100 AU for a given planet mass. The cyan, yellow and orange lines correspond to cases for low, solar and high opacity. There are big differences in the survival trends for planet mass and metallicity between the two models: we see no correlation for quenched migration models compared to an anti-correlation in the super migration case.}
\label{fig:M_survival}
\end{figure*}

The metallic composition of protoplanets alters their collapse timescale. Fragments with higher metallicity are more opaque and therefore cool and collapse slower, while the opposite is true for low metallicity fragments \citep{HelledBodenheimer11}. This however, is true only if grain growth and settling are neglected. 

Figure \ref{fig:M_survival} plots the survival percentage of planets as a function of planet mass and metallicity, assuming that opacity scales with metallicity, for both the quenched and super migration cases. We ran 1000 discs for each planet mass at each metallicity, with disc parameters randomly sampled from the ranges shown in Table \ref{table:pop_vals}. The solid, dashed and dotted lines show planets surviving in 0-5, 5-20 \& 20-300 AU bins after disc dissipation. We binned the results in this way in order to compare with the radial velocity/transit and direct imaging surveys below.

In the quenched migration case we see that planets with masses greater than 4 $M_J$ are much more likely to survive in the 20-300 AU bin compared to the 0-5 bin. As we have already seen, quenched migration tends to strand gap-opening planets at wide orbits.
Additionally, in the quenched migration model there is almost no dependence of the survival percentage of protoplanets on the host star metallicity.
Most of the quenched migration protoplanets open gaps and so will have collapsed before they can be tidally disrupted. In this case the differences in the collapse timescale due to planet metallicity (ie Figure \ref{fig:allona_models2}) are too small to affect the results.

The trends are very different for the super migration case. We see that planets are much more likely to survive at small separations due to the reduced efficiency of gap opening. We also see a stronger metallicty dependence for the protoplanets surviving in the 0-5 AU bin. For instance, at $M_P = 5 M_J$ we see that [M/H]=-0.3 planets are more than four times as likely to survive than [M/H]=0.3 planets. These results corroborate and quantify the predictions made by \cite{HelledBodenheimer11}, but we have now seen that [M/H] is important only if the planet migration time scale is comparable to the contraction time.

There is some evidence that giant planet occurrence frequency is independent of metallicity above $M_P = 4 M_J$ \citep{SantosEtal17}, but there is no observational evidence to support such a strong anti-correlation. We elaborate more on this point in the discussion section.

\section{Models compared to observations}
\label{sec:obs_comp}

\begin{figure*}
\includegraphics[width=1.8\columnwidth]{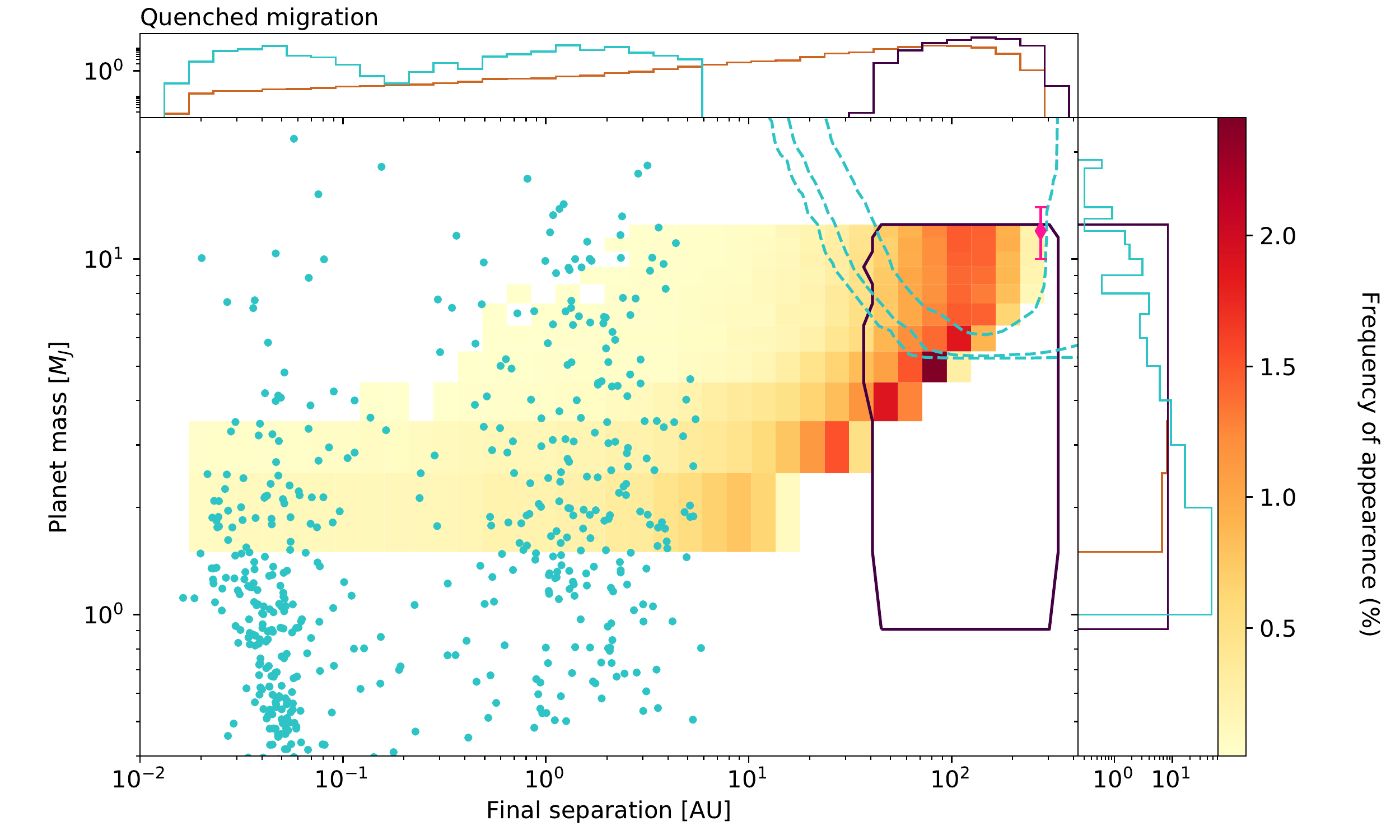}
\includegraphics[width=1.8\columnwidth]{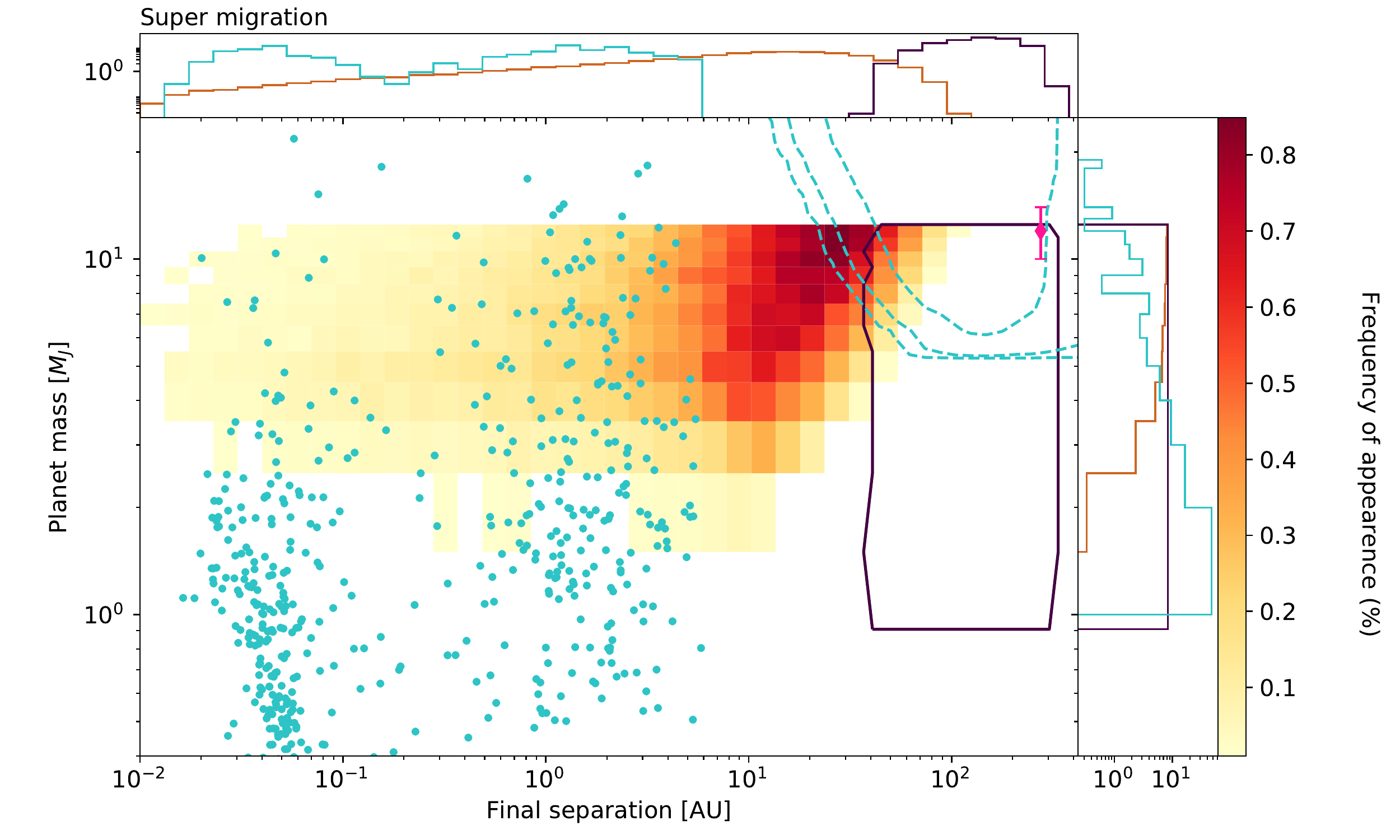}

\caption{Final probability distribution of the synthetic planets for the quenched (top) and super (bottom) migration models (orange-red colour map). The initial population of planets is indicated by the dark purple contour. Observed data for planets around stars of mass 0.7-1.4$M_{\odot}$ from exoplanets.org are plotted in turquoise. 
The dashed cyan lines plot the 0.5, 5 \& 50 \% completeness limits of the NaCo-LP direct imaging programme from \protect\cite{ViganEtal17} while the pink point marks the DI detection of AB Pic. 
Planets that survive in the quenched model are mainly stranded at wide orbits. In the super migration model, low mass planets are more likely to be tidally disrupted and planets tend to migrate much closer to the star. Additionally, the majority of super migration planets are not well sampled by the DI observations.}
\label{fig:a_MP_plot_Q}
\label{fig:a_MP_plot_S}
\end{figure*}

\subsection{Synthetic planets}\label{sec:PS-results}

In order to compare our models with observations we ran 10,000 simulations for each planet mass [0.5, 1, 2, ...,  12~$\mj$] in discs with parameters randomly chosen from the ranges in Table \ref{table:pop_vals}\footnote{The dependence on sample size is examined in Appendix \ref{sec:app_KS}.}. This corresponds to an initial planet mass function $N(M_P) \propto M_P^{0}$ (we later consider a $N(M_P) \propto M_P^{-1.3}$ case in Section \ref{sec:models-vs-obs-detail} and Appendix \ref{App:steep_imf}). We show the final mass-separation distribution of planets that survive the migration and disc dissipation phase for the quenched and super migration scenarios in Figure \ref{fig:a_MP_plot_Q}.  

For comparison, we have also plotted as cyan dots the planets from exoplanets.org for systems with stars in the mass range 0.7-1.4 $M_{\odot}$. \footnote{While we do not use the exoplants.org data directly in this paper, this data serve to indicate how far our synthetic planets need to migrate.}
The dark purple contour at 80-300 AU marks the starting location of our initial planet population. The orange-red colour maps show the percentage of the initial planet population that survive disc dissipation in a given separation and mass bin of the diagram. The dashed cyan lines represent the 0.5, 5 \& 50\% completeness limits of the NaCo-LP DI survey presented in \cite{ViganEtal17}.
The histograms above and to the right of the main plot show the separation and the mass distributions of the planets respectively. The dark purple and orange lines correspond to the initial and final populations of the synthetic planets.

First consider upper panel of Figure \ref{fig:a_MP_plot_Q} for the quenched migration case. We see that the efficient gap opening prescription has left many gas giants stranded at wide orbits beyond 50 AU. Only a few planets seem to reach the inner disc. The large fraction of planets left at tens of AU is contradicted by the direct imaging results showing that wide orbit gas giants are rare. Therefore, in the quenched migration scenario the fragmentation of massive discs into planetary mass clumps must be very rare, in agreement with the results of \cite{ViganEtal17}.  
In contrast, the lower panel of Figure \ref{fig:a_MP_plot_S} shows that the peak of the planet separation distribution is much closer to the star for the super migration model, it is still hard to detect these planets with current DI surveys.
The hatching of planetary mass clumps by GI may therefore be more prevalent in the super migration scenario, as concluded by previous authors \citep{Nayakshin17a,MullerEtal18}. 

The two migration extremes differ not only in the separation distribution of the survived planets but also in the mass function of these planets at different separations. 
The quenched migration mass function inside 20 AU comprises of mostly $\sim 2-3\mj$ mass planets whereas the mass function beyond 20 AU is dominated by the most massive planets. This is because only the low mass planets are able to migrate deep inside the disc in this scenario. The lack of planets below $2\mj$ is due to these planets being tidally disrupted (recall that these planets are below the contraction time curve in the left panel of Figure \ref{fig:M_transition}). The corollary of this is that high mass planets are very rarely able to migrate deep into the disc in the quenched migration model, and hence dominate the wide orbit spectrum.

Super migration has a qualitatively similar mass function with separation, 
though planets are mostly clustered between 5 and 30 AU. Massive ($M_P \gtrsim 4-6$~$\mj$) gas giants dominate the mass function beyond 20 AU but are unable to efficiently migrate to smaller separations. 
The minimum planet mass has now risen to $3-4 \mj$. Less massive gas giants in this scenario migrate inwards too rapidly and are tidally disrupted (being below the dashed curve in the right panel of Figure \ref{fig:M_transition}).
A more detailed examination of these mass functions can be seen later in Figure \ref{fig:M_hists}.

\subsection{A brief observational summary}\label{sec:obs_sum}

Recent observational results may allow us to better constrain GI planet formation models, here we present a summary of the most relevant findings.

\begin{itemize}
   \item \cite{CummingEtal08} used radial velocity measurements to constrain the occurrence rate of planets in the mass range 2-15 $\mj$ inside of 5 AU to be 4.2\%. This limits the number of GI planets that can survive migration from wide separations inside 5 AU.

    \item Gas giants and brown dwarfs are very rare at wide separations. Using direct imaging, \cite{ViganEtal17} found an occurrence rate of 2.1\% for 0.5-75 $M_J$ planets around FGK stars in the range 20-300 AU\footnote{This result is also supported by the recent GPIES DI survey for their FGK sample \citep{NielsonEtal19}}. The only DI detection in the planetary regime at a wide-orbit around an FGK star is AB Pic, a 10-14 $M_J$ object at 275 AU \citep{ChauvinEtal05,BonnefoyEtal10}. It is difficult to interpret this as a primordial formation frequency for GI planets however due to uncertainties in migration, gap-opening, planet-planet interactions and gas accretion immediately after the disc fragmentation phase. Furthermore, these limits assume wide orbit planets formed according to hot start models.

    \item There is a growing evidence for a population of massive gas giants and brown dwarfs at separations of less than a few AU with properties consistent with the GI formation mechanism. The mass above which GI formation may dominate over formation via Core Accretion is disputed, and may be as low as $\sim 1-2 \mj$ \citep{SuzukiEtal18,MaldonadoEtal19} or as high as $4-10 \mj$ \citep{SantosEtal17,Schlaufman18}. At higher masses, \cite{TroupEtal16} found no correlation of brown dwarf occurrence rate with host star metallicity, in contradiction to the predictions of CA models \citep{MordasiniEtal12}. 
    
\end{itemize}

\subsection{Models vs observations}\label{sec:models-vs-obs-detail}
\begin{table}
\centering
\begin{tabular}{c|ccc}
             &$M_P$ &  $F_{\rm best}$ & [$F_{\rm min}$,$F_{\rm max}$]    \\
             &      &  (\%)     & CL=95\% \\
\hline

\bf{Quenched}                   & <12 $M_J$  &                 &                \\ 
$N(M_P) \propto M_P^{0}$           &         & 2.7             & 0.7, 14.7      \\ 
$N(M_P) \propto M_P^{-1.3}$        &         & 3.3             & 0.8, 17.8        \\ 
\hline

\bf{Super}                      & <12 $M_J$ &                 &                \\
$N(M_P) \propto M_P^{0}$           &         & 7.4            & 1.8, 40.1      \\ 
$N(M_P) \propto M_P^{-1.3}$        &         & 33.0            & 7.2, 95.6      \\ 
\hline
\end{tabular}
\caption{This table combines our quenched and super migration models with DI observational constraints from \protect\cite{ViganEtal17}. $F_{\rm best}$ is a percentage for the number of systems that could have undergone one fragmentation and still be consistent with the DI observations, $F_{\rm min}$ and $F_{\rm max}$ are the corresponding 95\% confidence levels. The results are presented for both flat and steep initial mass functions. We see that quenched migration is consistent with GI occurring in only a few percent of systems, whilst super migration allows GI to occur much more frequently. The sensitivity of current DI surveys becomes increasing poor below 10 $M_J$, an initial mass function that peaks at around 1 or 2 $M_J$ allows many more planets to `hide' below this sensitivity limit.}
\label{table:occurrence_rate}
\end{table}

A successful model needs to explain all three of the observational issues noted in the previous section. We now argue that the super migration scenario comes closest in terms of reproducing the populations inside 5 AU and outside 20 AU within our model, but we caution that both migration regimes must be examined further with physics not yet included in our models, such as gas and solid  accretion, and core formation inside the fragments.

In comparing our models to the direct imaging constraints, we note that the exact observational limits on our population synthesis results depends on  a non-trivial modelling that takes into account observational biases for the specific synthetic distribution. 
In order to take this in to account in the best possible way, we adopted an approach similar to the one described in \cite{ViganEtal17} and used the QMESS Monte Carlo simulation code \citep{BonavitaEtal13} to couple the information on the detection probability from the NaCo-LP extended survey \citep{ViganEtal17} with the populations described in Section~\ref{sec:PS-results}. 
QMESS applies a re-scaling of the detection probability using the planet survival rates from both our quenched and super migration models, which allows us to test the detectability of the planets in the synthetic populations, given the sensitivity limits from the specific survey. 
Using the method described by \cite{lafreniere07}, we then calculated the value of the frequency of systems that could undergo GI with one fragment and still be consistent with the DI observations and summarise this in  Table~\ref{table:occurrence_rate}.
The large error bars on the frequency are a consequence of the small number of DI detections. Only one target in the sample considered by \cite{ViganEtal17} has a companion with mass lower than 12~$M_J$ and separation between 20-300 AU (AB~Pic, see \cite{ChauvinEtal05} and \cite{BonnefoyEtal10} for details). 
Moreover, as shown in Figure~\ref{fig:a_MP_plot_Q}, the DI observations cover a small portion of the parameter space populated by our GI models and the only DI detection falls in a region where the survival rate is low, especially in the case of the super migration. 
This causes the frequency posterior distribution to be biased towards higher values, therefore  the values of $F_{best}$ reported in Table~\ref{table:occurrence_rate} should be interpreted as upper limits. Especially for the super migration model, the majority of the final population is hidden below the DI sensitivity limits.

\begin{figure*}
\includegraphics[width=1.0\columnwidth]{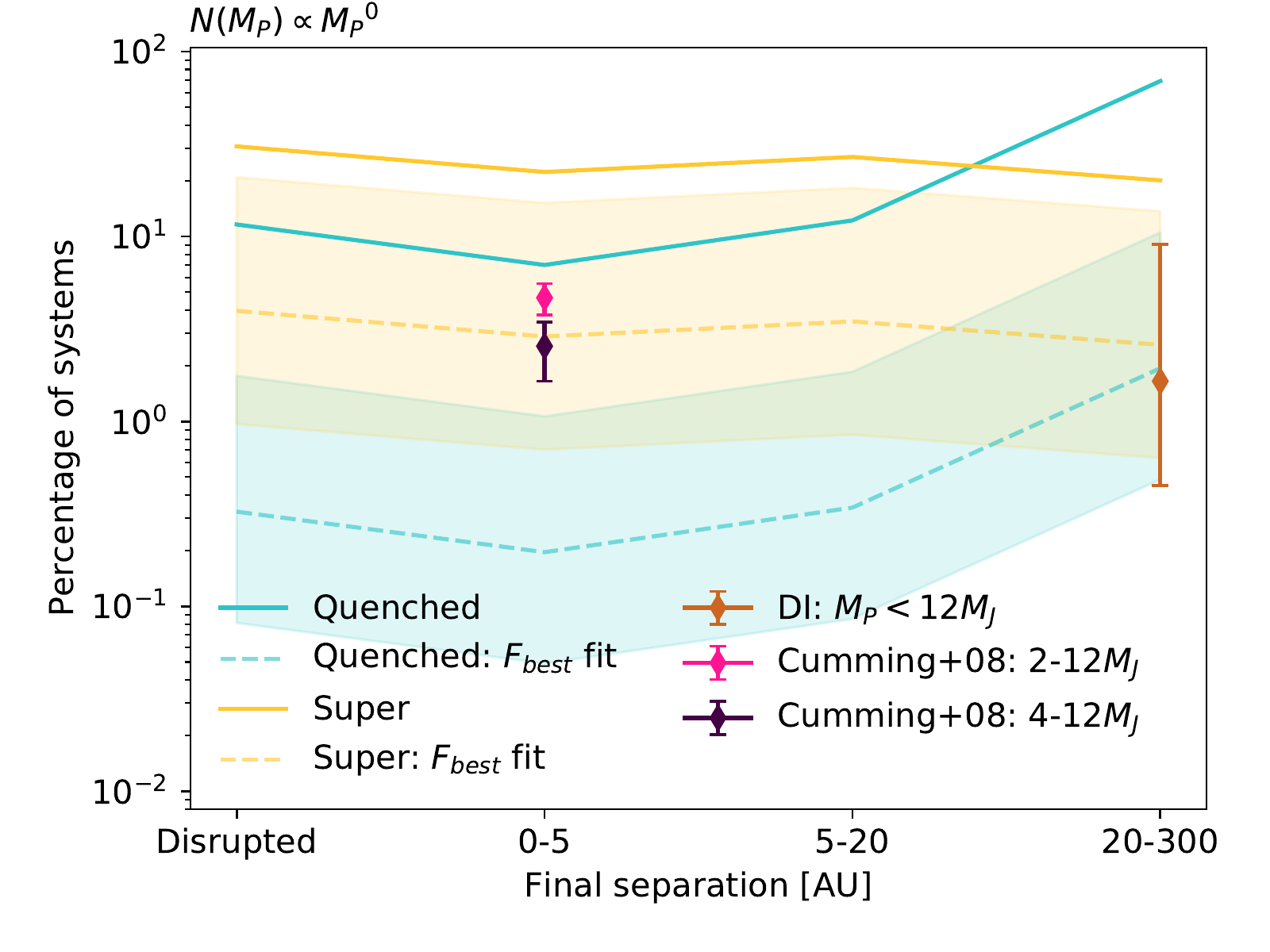}
\includegraphics[width=1.0\columnwidth]{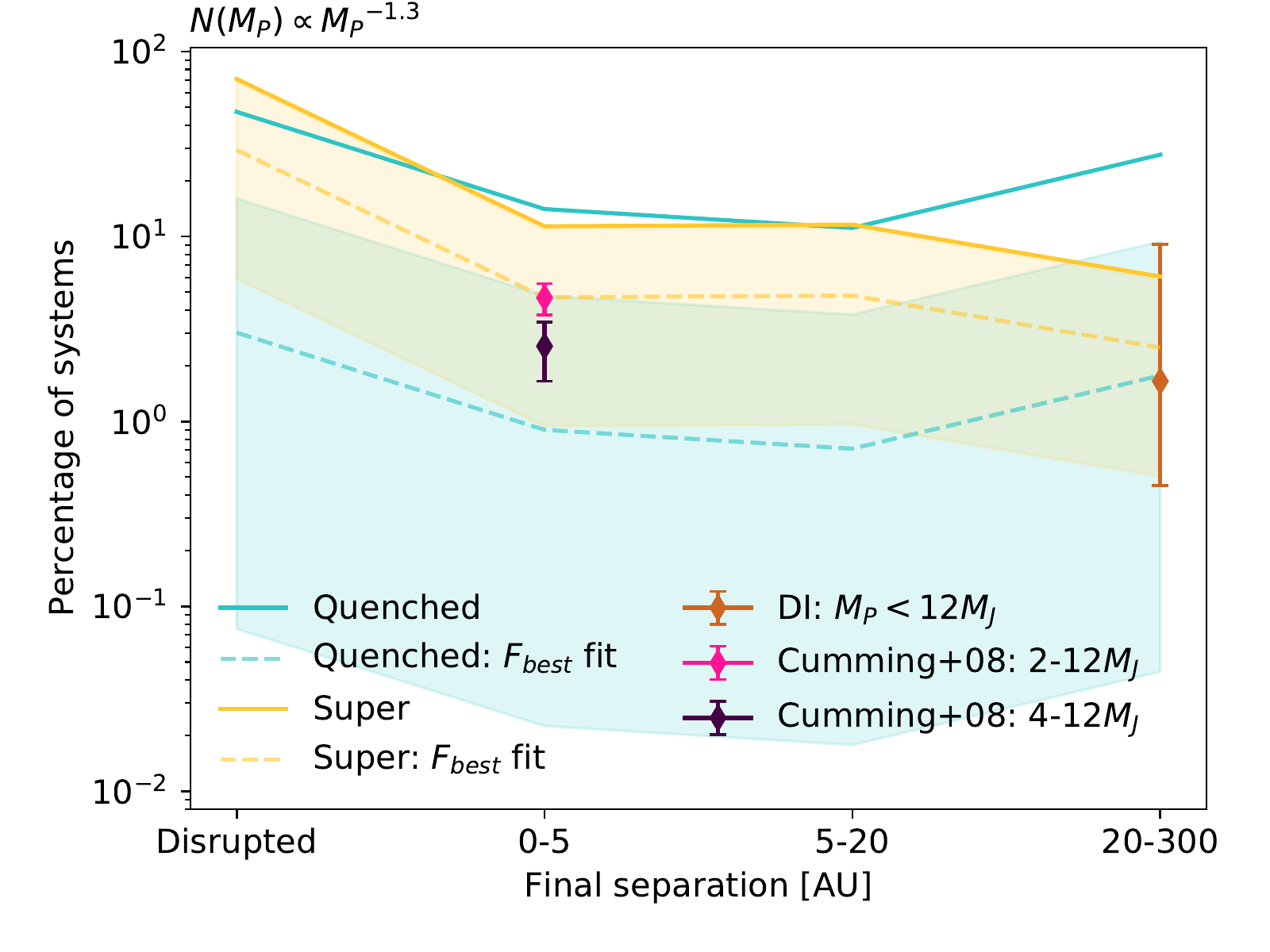}
\caption{Final population predictions for the quenched (blue) and super (yellow) migration models with 1000 runs for each planet mass in the range [1-12] $M_J$. These plots also include observational constraints: in orange we show wide orbit DI constraints recalculated for our mass ranges using the \protect\cite{ViganEtal17} sample. In pink and purple we plot the detection statistics for 2-12 and 4-12 $M_J$ planets inside 5 AU from the KECK survey \citep{CummingEtal08}. 
The dashed lines show the $F_{best}$ frequencies that are consistent with the DI observations, as given in Table \ref{table:occurrence_rate}. The shaded area shows the 95\% confidence levels.
Different initial mass functions for the protoplanets are shown for $N(M_P) \propto M_P^{0}$ (left) and $N(M_P) \propto M_P^{-1.3}$ (right). 
Both super migration models and the quenched model with a steep initial mass function are consistent with explaining 100\% of 4-12 $M_J$ planets inside 5 AU.
}
\label{fig:Pop_N_Mpf}
\end{figure*}

The solid lines in Figure \ref{fig:Pop_N_Mpf} show the percentage of planets in the models from Figure \ref{fig:a_MP_plot_Q} that were either tidally disrupted (the first bin) or survived at separations of 0-5, 5-20 and 20-300 AU. For each solid curve in Figure \ref{fig:Pop_N_Mpf} the sum of the four bins adds to 100\%. The dashed curves are the same as the solid curves of the same colour, but multiplied by the corresponding value of $F_{\rm best}$ from Table \ref{table:occurrence_rate}. The dashed curves have therefore been re-scaled to be consistent with the DI observations. The shaded areas represent the 95\% confidence intervals on these values; these are very wide due to the low number of detections in the current DI surveys. 
In the left panel we show the results for an initial planet mass spectrum of $N(M_P) \propto M_P^{0}$, while in the right one we show the values for a spectrum $N(M_P) \propto M_P^{-1.3}$.

We have added two additional sets of observational constraints to these figures. In the 20-300 AU bin we mark the DI observational constraints from \cite{ViganEtal17}, adjusted slightly for planets in the mass range 1-12$M_J$. As described above, these constraints coupled with the Quick-MESS code gave us the $F_{best}$ values used to plot the dashed shaded areas. We have also added observational constraints in the 0-5 bin from the radial velocity (RV) detections by the KECK telescope \citep{CummingEtal08}. These provide an indicator of the number of gas giants in the mass ranges 2-12 and 4-12 at separations less than 5 AU. 

First consider the left part of Figure \ref{fig:Pop_N_Mpf}. We see that for a flat initial mass spectrum the RV constraints on gas giants inside 5 AU are well fitted by the $F_{\rm best}$ super migration model. Super migration is able to explain the full population of super Jupiter gas giants inside 5 AU whilst simultaneously being consistent with a lack of planets at wide orbits. By comparison, for a flat initial mass spectrum quenched migration can only explain around 10\% of the RV giants below 5 AU. 

On the right hand side we plot the results for a steeper planet mass function. We chose the power -1.3 to be consistent with the mass function found by \cite{CummingEtal08} for giants inside 5 AU. It is not clear that these two functions should be the same, but we make this choice to avoid being arbitrary and also to demonstrate the qualitative effect of a steep initial mass function.  With the steep mass function, the adjusted $F_{\rm best}$ curves for both quenched and super migration population synthesis are consistent with the total numbers of planets inside 5 AU from RV measurements.

\subsection{A closer look at the planet mass functions}\label{sec:mass_function}

\begin{figure*}
\includegraphics[width=1.0\columnwidth]{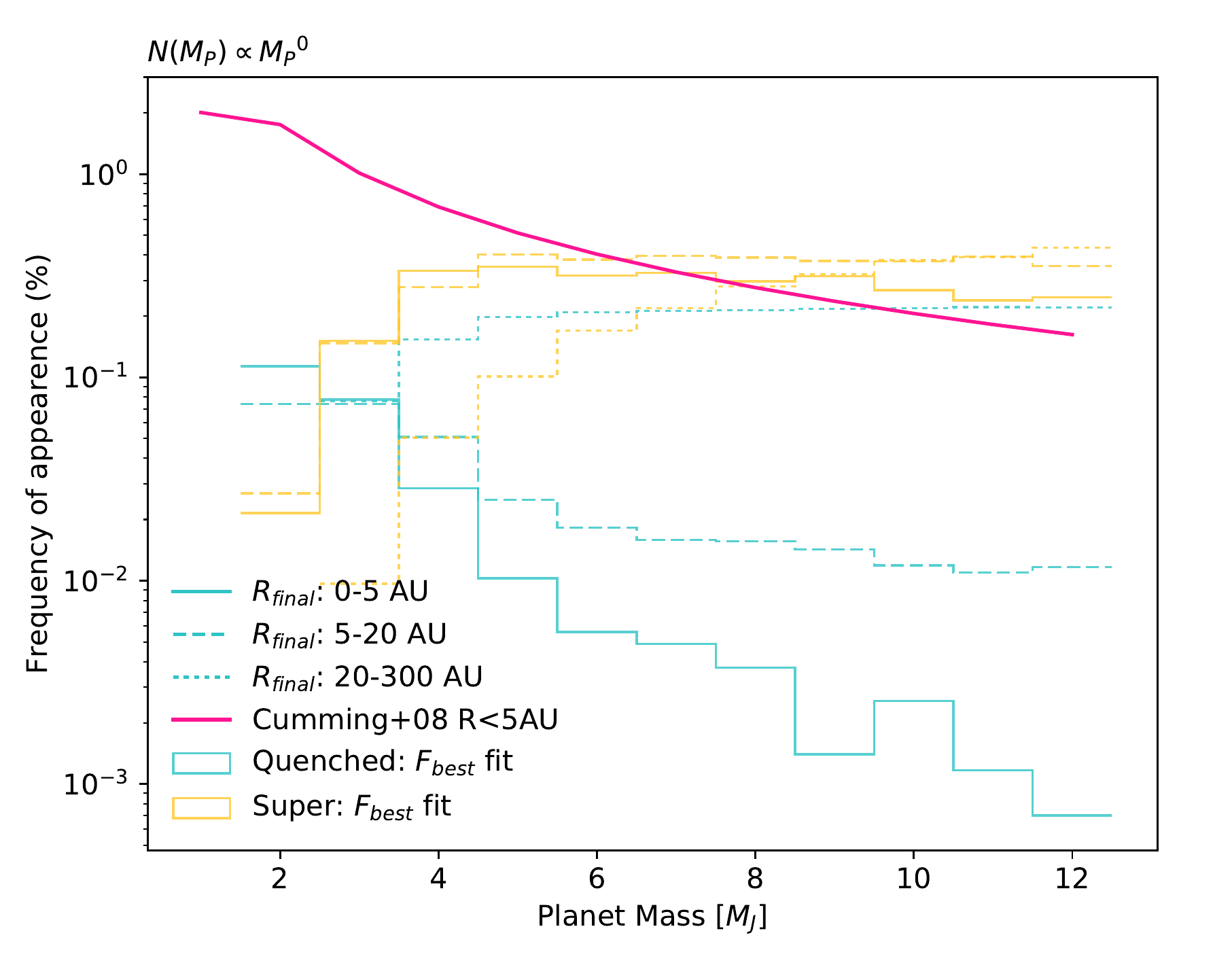}
\includegraphics[width=1.0\columnwidth]{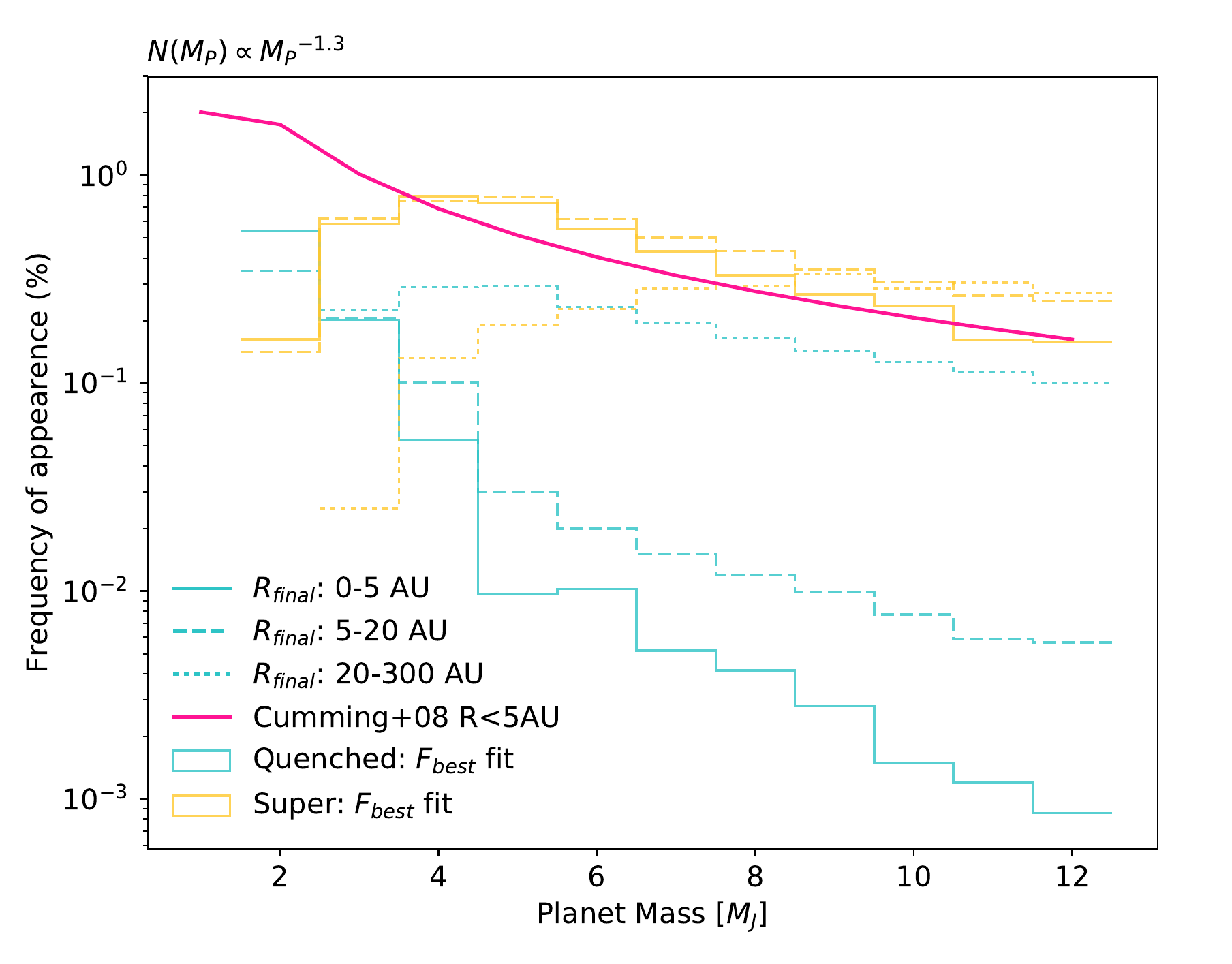}
\caption{Final mass functions for the quenched (blue) and super (yellow) migration models with 1000 runs for each planet mass in the range [1-12] $M_J$, adjusted to the $F_{\rm best}$ fit frequencies. The initial mass functions are $N(M_P) \propto M_P^{0}$ (left) and $N(M_P) \propto M_P^{-1.3}$ (right). Solid, dashed and dotted lines show the separation bins of 0-5, 5-20 \& 20-300 AU, respectively. The pink line plots the mass function inside 5 AU from RV measurements from \protect\cite{CummingEtal08}. Looking at the solid lines, both super migration models provide a good fit to the data above 4 $M_J$, whilst the quenched migration models underpredict the number of planets inside 5 AU by an order of magnitude.}
\label{fig:M_hists}
\end{figure*}

In the previous section we showed that  both migration scenarios can yield a reasonable explanation of the data in terms of the RV and DI planet occurrence frequencies, in particular for the steeper initial planet mass spectrum. In the final part of this analysis we consider the planet mass function {\em in different radial bins} and compare them to the \cite{CummingEtal08} mass spectrum for RV planets. Figure \ref{fig:M_hists} shows the mass function for the $F_{\rm best}$ adjusted quenched and super migration models in each of the three separation bins from the previous section: 0-5 (solid), 5-20 (dashed) and 20-300 AU (dotted). The left and right panels show the case for our two choices of the initial mass function. On each figure we have also plotted the \cite{CummingEtal08} RV mass function for gas giants inside 5 AU.

Focusing on the solid curves, we see that the quenched migration model deposits too few planets inside 5 AU. The super migration model is much closer, but still slightly flatter than the \cite{CummingEtal08} spectrum. We emphasise that the quenched migration result cannot be improved by rescaling the curve up. This would require more initial clumps and hence violate the DI constraints.

The right panel of Figure \ref{fig:M_hists} shows that for the steeper initial mass function we find another negative result for the quenched migration, but a more promising one for the super migration. Whilst the quenched migration model again massively underpredicts the occurrence frequency of planets inside 5 AU, the super migration model agrees well with the RV spectrum above 4 $M_J$, both in terms of normalisation and shape.

\section{Discussion}
\label{Sec:Discussion}

\subsection{Main results of the paper}\label{sec:Discussion_results}

Our study suggests that the survival of GI planets at all separations depends strongly on their ability to open gaps before they reach $\sim$ 20 AU. Protoplanets that do not open gaps are tidally destroyed: we saw in Figure \ref{fig:M_transition} that the type I migration timescale is always shorter than the planet collapse timescale. If a protoplanet manages to open a gap, its migration slows down and it may have enough time to cool and collapse into a dense post-collapse GI planet. This implies that only GI planets that are able to open gaps in the disc outside of $\sim$ 20 AU can survive tidal disruption. While the mechanics of the gap opening process itself is now relatively well understood \citep[e.g.,][]{MalikEtal15}, what is appropriate for rapidly migrating giant planets in massive discs is an open question since the exact parameters of the discs at fragmentation remain poorly constrained \citep{KratterL16}. This is one of the main uncertainties of the GI theory of planet formation. 

In this paper we tried a different tack in order to make progress. By considering both the directly imaged wide-orbit constraints and radial velocity observations of massive gas giants at separations inside 5 AU we hoped to put some bounds on the efficiency of planet migration. We studied two extreme cases for the gap opening uncertainty that we termed quenched and super migration (Section \ref{sec:gap_criteria}). Gap opening is efficient in quenched migration, protoplanets open gaps easily and are frequently stranded in the outer disc. The inverse is true for super migration (see Section \ref{sec:gap_imp}).

We used the QMESS \citep{BonavitaEtal13} code to compute $F_{\rm best}$, the planet occurrence rate at wide separations, for both models under the constraints of the NaCo-LP DI survey \citep{ViganEtal17}. This analysis (Section \ref{sec:models-vs-obs-detail} and Table \ref{table:occurrence_rate}) gives us $F_{\rm best}$ occurrence frequencies of 2.7\% and 3.3\% for quenched migration for the the initial mass functions $dN/d M_P \propto M_P^0$ and $dN/d M_P \propto M_P^{-1.3}$, respectively. The latter mass function corresponds to the one derived by \cite{CummingEtal08} for RV gas giants. Our results for the quenched migration models are comparable to those of \cite{ViganEtal17} who found that GI only occurred in 1-8.6\% of systems. This is to be expected since the gap opening criteria of the underlying population synthesis for quenched migration is similar to those used by \cite{ViganEtal17}.
For the super migration scenario, we derived values of 7.4\% and 33.0\% for $F_{\rm best}$ for the flat and the steep initial mass functions respectively. These values are higher than $F_{\rm best}$ for the quenched migration population since super migration transfers wide orbit protoplanets into the inner disc more efficiently, and also destroys them by tidal disruption more often.
However, due to the lack of DI detections in the wide-orbit gas giant parameter space, it is very hard to constrain these values. In fact, the upper 95\% confidence limits for the two quenched migration models are 14.7\% and 17.8\%, rising to 40.1\% and even 95.6\% for super migration. Future models will need to explore the mass parameter space in a continuous way from Jupiter through to stellar mass in order to properly use the full set of DI observations.

Using the $F_{\rm best}$ direct imaging constraints we demonstrated in Figure \ref{fig:Pop_N_Mpf} that super migration models are able to simultaneously explain both the small wide orbit population and the larger population of gas giants inside 5 AU found via RV surveys for both the flat and the steep initial protoplant mass functions. By comparison, the quenched migration model underpredicted the number of planets inside 5 AU (although less strongly for the steep mass function). 
Figure \ref{fig:M_hists} shows the mass function of each model. The super migration models provide reasonable fits to the mass function for 4-12 $M_J$ planets whilst the quenched migration models massively underpredict the number of planets above 4 $M_J$ by over an order of magnitude. The tidal disruption of low mass planets provides an interesting prediction: only GI planets more massive than $\sim$ 2-4 $\mj$ are able to cool and collapse fast enough to avoid tidal disruption, regardless of the minimum mass for protoplanets born at very wide orbits.

Our results are significant with respect to the turnover in the correlation of host star metallicity with gas giant frequency in the range 2-10 $\mj$ \citep{SantosEtal17,Schlaufman18,NarangEtal18,Adibekyan19,MaldonadoEtal19}. These observations suggest that planet formation is dominated by GI above this critical mass, since CA models predict a strengthening of the metallicity relation for massive gas giants and brown dwarfs \citep{MordasiniEtal12}. This is a severe problem for our quenched migration models since they fail to produce enough planets inside 5 AU whilst simultaneously obeying the DI constraints. As seen in Figure \ref{fig:M_hists}, quenched migration models underpredict the occurence rate of 4 $\mj$ gas giants inside 5 AU by an order of magnitude, rising to two orders of magnitude for 10 $\mj$ giants. We find that the quenched migration model of GI with a low occurence frequency is incompatible with explaining the population of massive gas giants inside 5 AU, and therefore cannot explain the turnover in metallicity correlation.

We find no correlation between the planet survival frequency and host star metallicity in the quenched migration model. This is in agreement with \cite{SantosEtal17} who found no correlation of gas giant frequency with host star metallicity for planets more massive than 4 $M_J$. 
In the super migration model, we see a negative correlation of planet frequency with host metallicity (Figure  \ref{fig:M_survival}), as predicted by \cite{HelledBodenheimer11}. There is a factor $\sim 2$ reduction in the planet occurrence rate for a factor 4 increase in host star metal abundance for a flat initial mass spectrum. In observational terms this is rather a weak correlation, especially compared to the strong \cite{FischerValenti05} correlation for less massive gas giants which goes as $N(M_P) \propto 10^{2 [M/H]}$.
These results therefore update the \cite{HelledBodenheimer11} prediction: the occurrence rate of GI planets will only negatively correlate with host star metallicity if the migration timescale is comparable to the protoplanet cooling timescale.
However, it is probably a mistake to infer too much from these correlations. In this study we have assumed that protoplanets have a fixed metallicity and neglected the effects of heavy-element  accretion and core formation. We know that if $\sim$ 1\% of the disc mass is in the form of planetesimals/pebbles then the metallic composition of wide orbit gas giants can be enhanced considerably by planetesimal and/or pebble accretion \citep{HelledEtal06,HelledEtal08,HumphriesNayakshin18}. How these pebbles affect the metallicity correlations will be a subject for future studies.

Based on these results, we suggest that rapid inward migration of protoplanets born in the outer disc must be a significant part of the reason why wide orbit gas giants are so rare. 
If all massive gas giants above a few $\mj$ inside 5 AU are formed by GI (as suggested by metallicity correlations), then our best fitting scenario is a super migration model with a steep initial mass function in which GI protoplanets form in at least tens of percent of systems. Reducing this fraction means that our models underpredict the occurence rate of sub 5 AU massive gas giants (as seen in Figure \ref{fig:M_hists}). 

Further observations are necessary in order to better constrain the turnover mass for gas giant metallicity correlations, but if GI gas giants contribute significantly to this population then we have shown that protoplanet formation via GI must occur more frequently than previously thought.

There remain a number of physical processes to be included in future work (see the following section), all of which are likely to increase this estimate further since they act to transform protoplanets of a few Jupiter mass into something else (e.g., a sub-Neptune planet or a stellar mass companion).

\subsection{Missing physics}\label{sec:missing}

\subsubsection{Gas accretion onto clumps}
\label{sec:discussion_gasacc}

Gas accretion rates onto protoplanets depend heavily on the cooling rate of the local disc gas \citep{Nayakshin17a,HumphriesNayakshin18}. Essentially, gas must be able to cool faster than it is swept through the Hill sphere of the planet by the Keplerian shear in the disc. Gas accretion rates are therefore highly dependent on the opacity of accreted gas, which is set by its dust content. We can see from Figures \ref{fig:allona_models} and \ref{fig:allona_models2} that more massive protoplanets cool faster due to their higher luminosities. Shorter cooling timescales imply that they can accrete gas from the disc and grow in mass at a higher rate. This may propel them into the brown dwarf mass-regime and beyond. 

It is possible that including gas accretion could relieve the issue that both models have with an overabundance of gas giants above $\sim$10$M_J$, stretching the high mass population to larger masses may remove the apparent bias towards higher mass planets. This would act in a degenerate way with steepening the initial mass function for fragmentation.
Gravitational instability has often been invoked as a mechanism for forming brown dwarfs \citep{KratterEtal10,StamatellosInutsuka18,MoeEtal18} though this has not yet been implemented in a universal population synthesis model. These planets would need to undergo a runaway accretion phase since we know the occurrence rate of 2-75$M_J$ objects is rare at wide separations \citep{ViganEtal17}. Gas accretion would need to propel these objects into the companion mass regime where there is a peak at 200 $M_J$ at 40 AU \citep{DuquennoyMayor91, RaghavanEtal10}. Investigating gas accretion is very important since it will allow us to connect more closely with the DI observational constraints.

\subsubsection{Heavy-element accretion,  core formation, and opacity/metallicity scaling}
\label{sec:pebs}

In this work we have also neglected the role of planetesimal and/or pebble accretion and core formation on the evolution of GI protoplanets. From the results and discussion in this paper so far we have seen that in this limit GI can occur only rarely in real systems. However, if dust grains can grow to mm sizes during the self-gravity phase of young discs then we need to account for these additional processes.

In brief, both planetesimal and pebble capture by GI fragments is very efficient, providing mm pebbles exist in the disc at $10^4-10^5$ years \citep{HelledEtal06,HelledEtal08, Boley09, HumphriesNayakshin18}. Once captured, small grains may grow and rapidly sediment to form a core \citep{HelledEtal08,HS08}. The outcome is then uncertain. Grains also increase the opacity of fragments, increasing their collapse time. However, they also increase the fragment density which may in fact shorten their collapse time due to the impact it has on the hydrostatic equilibrium of the protoplanet. The interested reader should consult \cite{NayakshinFletcher15} for an analytical understanding of how these processes may affect the planet formation process in GI. 
Additionally, liberated gravitational potential energy during core formation may act to expand and prematurely disrupt fragments. This has been found analytically in \cite{Nayakshin16a} and demonstrated in 3D simulations (Humphries \& Nayakshin 2019 in prep.).
We leave more detailed study on the effects of pebble and planetesimal accretion and core formation to future work with the expectation that it may be crucial in determining the expected outcomes of the GI theory \citep{HelledEtalPP62014}.

Finally, in this work we assumed that the metallicity of the protoplanets scales with the assumed metallicity. In that case, more metal-rich objects have longer contraction timescales as they radiative cooling is inefficient. This trend, however, could reverse if grain growth and settling are included since this effect can significantly decrease the atmospheric opacity, and in that case all metallicities can lead to relatively short contraction timescales \citep{HelledBodenheimer11}. It is therefore desirable to investigate in future studies the relation between planetary metallicity and (grain) opacity, as well as the relation between planetary and stellar metallicity.

\subsubsection{Core Accretion planets}\label{sec:CA}

In this paper we have made the commonly adopted assumption that all wide separation giant planets are formed via GI. However, the Core Accretion scenario may also produce wide separation gas giants by, e.g., planet-planet scattering of smaller (sub ten AU) separation planets onto larger orbits \citep{MarleauEtal19}. If these planets evolve according to `cold start' models \citep{MarleyEtal07}, they are unlikely to be observed by the current generation of DI surveys and therefore will have no impact on our analysis.

Recent comparisons of the luminosity of giant planets with sets of evolutionary tracks or formation models (e.g. \citep{JansonEtal11, BonnefoyEtal13} Janson et al. 2011; Bonnefoy et al. 2013) tend to show that they do not appear to be compatible with very low initial entropy, cold start models but instead with  intermediate or high initial entropy warm and hot start models. This has been further confirmed by the latest dynamical constraints for the masses of directly imaged companions \citep{RodetEtal18, SnellenBrown18, CalissendorffJanson18,FlaggEtal19}.

In fact, in both formation scenarios the primordial entropy (cold/warm/hot start) of the planets remains uncertain. Various studies have showed that even CA planets may `warm start' \citep{Mordasini13, BerardoEtal17, CummingEtal18} whilst GI planets may have different primordial states when accounting for more complex physics such as core formation, accretion, rotation, opacity evolution, etc. In this current literature, both formation models may lead to a range of luminosities, and this topic should be investigated in more detail in order to allow a more complete interpretation of direct imaging surveys\footnote{However, for the NaCo-LP survey data that we use in this paper many stars are over $10^8$ years old, by which point most evolutionary models have converged anyway.}.

\section{Conclusions}
\label{Sec:Conclusions}
In this work we investigated a fixed planet mass population synthesis model for protoplanets born through gravitational instability, coupled with state-of-the-art pre-collapse protoplanet evolution calculations.
We found that the survival of GI planets is highly dependent on their ability to open gaps in the disc outside of $\sim$ 20 AU. Without gap opening, migration is always faster than the protoplanet collapse timescale and so protoplanets will be tidally disrupted. Due to the balance of migration and collapse times we found that the minimum planet mass able to survive tidal disruption was $\sim$ 3 $M_J$. It is therefore very important that future studies of gravitational instability consider the timescale on which GI protoplanets collapse to become the compact planets that are observed at late times.

We also compared our results to observations: DI constraints on the frequency of wide orbit companions \citep{ViganEtal17}, RV constraints on gas giants inside 5 AU \citep{CummingEtal08}, and correlations with host metallicity from \cite{FischerValenti05}, \cite{SantosEtal17} and \cite{Schlaufman18}. To assess the implications of these constraints we tested two extreme models for gap opening that we named quenched and super migration.

\begin{itemize}
    \item In the quenched migration model, planets efficiently open gaps and so are typically stranded in the outer disc. This violates the direct imaging constraints for gas giants beyond $\sim$ 20 AU \citep{ViganEtal17}. In order to obey this constraint, GI can only happen in a few percent of systems, depending on the initial protoplanet mass function. In this case, the quenched migration model can only form planets in the range 0-5 AU in less than 1\% of systems, more than a factor ten lower than the observed number. Therefore, under quenched migration GI planets may only provide a small contaminant to the population inside 5 AU. We found no correlation between the survival of planets and stellar metallicity for this model, in agreement with observations \citep{SantosEtal17,Schlaufman18}.

    \item For super migration, only a small percentage of planets are left at wide orbits which satisfies the direct imaging constraint. In this case up to several tens  of percent of systems can host a GI protoplanet, depending on the steepness of the initial mass function. Super migration provides a very good fit to the occurrence frequency of planets inside 5 AU and even provides a good match to the respective mass function at this separation. 
    This model reproduces the prediction by \cite{HelledBodenheimer11} that there should be a negative correlation between planet survival frequency and host star metallicity for GI planets, assuming that opacity correlates with metallicity, although the correlation is weak.
\end{itemize}

Quenched migration cannot simultaneously explain the dearth of planets at wide orbits and also the population of massive gas giants inside 5 AU. This is a problem since 
the change in metallicity correlation for gas giants in the range 2-10 $\mj$ implies that these objects may well have formed via GI \citep{SantosEtal17,Schlaufman18,MaldonadoEtal19}.
 By comparison, super migration can consistently explain both the lack of planets at wide orbits as well as the number and mass spectrum of giants inside 5 AU. If the super-Jupiter gas giant population inside 5 AU is dominated by GI planets then migration from the outer disc must be very efficient (see Figure \ref{fig:a_MP_plot_Q} for an overview of the final populations) and protoplanet formation via GI may well occur in at least tens of percent of systems.
 
This study should be considered as an additional step in the journey towards understanding planet formation in the gravitational instability scenario.
In future studies, we hope to include the additional physical processes of gas and solid accretion as well as core formation and feedback in order to explore how they change the characteristics of the final planetary population.

\section*{Acknowledgements}
JH and SN acknowledge support from STFC grants ST/N504117/1 and ST/N000757/1, as well as the STFC DiRAC HPC Facility (grant ST/H00856X/1 and ST/K000373/1). DiRAC is part of the National E-Infrastructure. MB gratefully acknowledges support from STFC grant ST/M001229/1 and RH acknowledges support from SNSF grant 200021\_169054. Part of this work was conducted within the framework of the National Centre for Competence in Research PlanetS, supported by the Swiss National Foundation. We would also like to acknowledge useful comments from the anonymous referee that helped to clarify some issues in this paper.


\bibliographystyle{mnras}
\bibliography{humphries}

\appendix
\section{Steeper initial mass functions}
\label{App:steep_imf}

\begin{figure*}
\includegraphics[width=1.8\columnwidth]{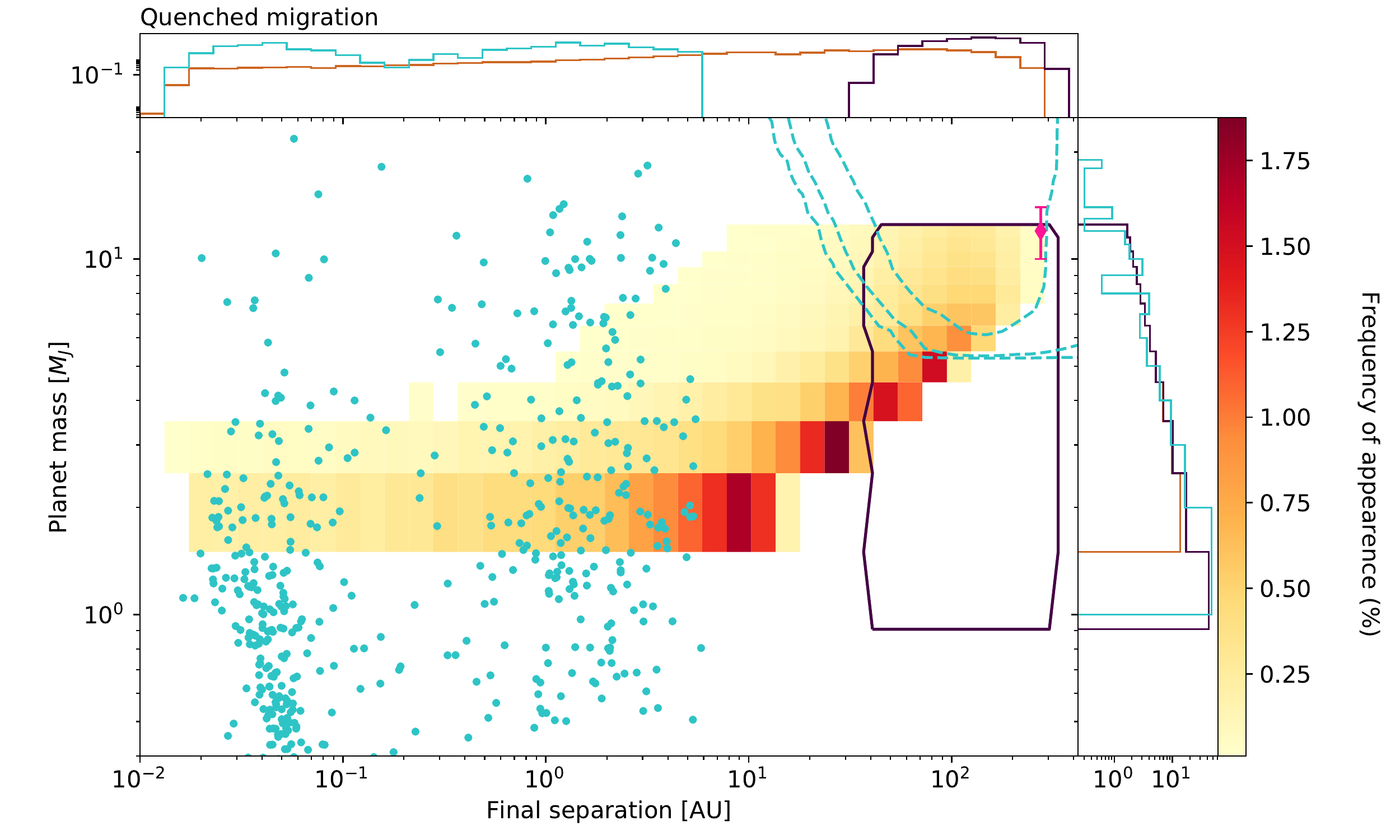}
\includegraphics[width=1.8\columnwidth]{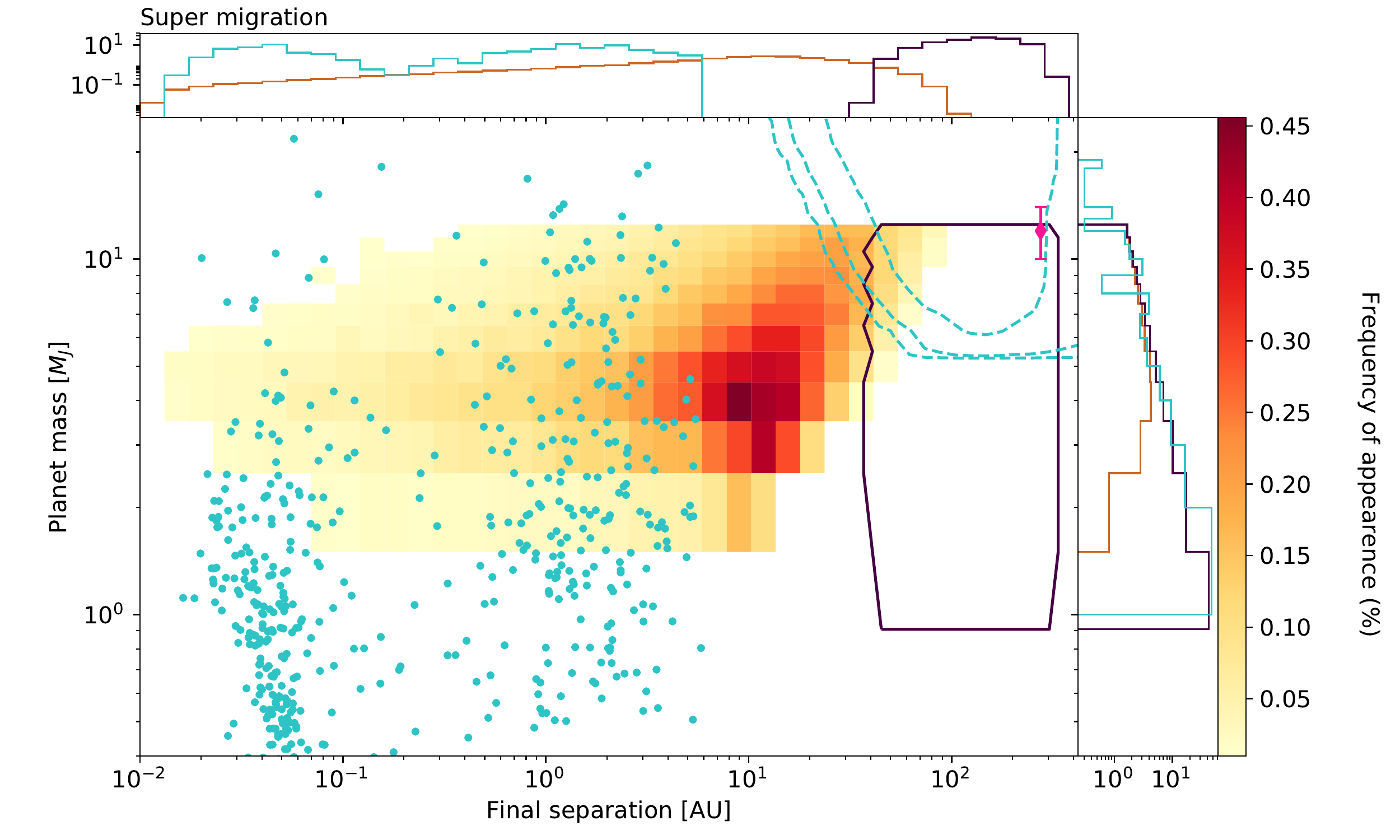}
\caption{Identical to Figure \ref{fig:a_MP_plot_Q} but for a steeper initial mass function of $N(M_P) \propto M_P^{-1.3}$. Notice that the number of massive planets has been reduced, moving the majority of the population into the poor sensitivity region of DI surveys.}
\label{fig:a_MP_plot_steep}
\end{figure*}

In this section we present Figure \ref{fig:a_MP_plot_steep}, a version of Figure \ref{fig:a_MP_plot_Q} for the steeper initial mass function of $N(M_P) \propto M_P^{-1.3}$. This initial mass function has reduced the number of high mass planets in each model. The sensitivity of current DI surveys becomes increasing poor below 10 $M_J$, an initial mass function that peaks at around 1 or 2 $M_J$ allows many more planets to `hide' below this sensitivity limit. 

\section{Sample size: Kolmogorov-Smirnov test}
\label{sec:app_KS}
When conducting population synthesis research, it is important to take a large enough sample such that the entire parameter space is well sampled. In Figure \ref{fig:a_MP_plot_Q} we chose 10,000 runs for each planet mass, giving us 120,000 sets of randomly selected disc parameters in total. In this appendix we outline the results of a two-dimensional, two-sample, Kolmogorov-Smirnov (KS) test to check that 120,000 is a sufficiently large sample size. We use an adaption of the test outlined in \cite{NumericalC07} by \cite{FasanoFranceschini87} to calculate this statistic.

The test will reject the null hypothesis that two samples were drawn from the same distribution if the final p value is less than the chosen significance value (for instance, $\alpha=0.01$). Additionally, approximations in this numerical algorithm mean that p values larger than 0.2 are inaccurate, but in any case such large p values indicate that the two populations are identical.

By comparing two randomly chosen parameter sets for the N=120,000 models, we obtained p values of 0.69 and 0.8 for the quenched and super migration cases. These large KS p values indicate that these populations are identical, there are no hidden islands of parameter space left unexplored. This is a reasonable conclusion due to the simplicity of our models and the degenerate affects of each parameter on the final outcome.





\bsp	
\label{lastpage}
\end{document}